\documentclass[]{JHEP}
\usepackage{epsfig}

\newcommand{\be}{\begin{equation}}
\newcommand{\ee}{\end{equation}}
\newcommand{\beqs}{\begin{eqnarray}}
\newcommand{\eeqs}{\end{eqnarray}}
\newcommand{\beqsn}{\begin{eqnarray*}}
\newcommand{\eeqsn}{\end{eqnarray*}}

\newcommand{\bea}{\begin{eqnarray}}
\newcommand{\eea}{\end{eqnarray}}
\expandafter\ifx\csname mathbbm\endcsname\relax

\title{Vacua, Propagators, and Holographic Probes in AdS/CFT}
\author{Ulf H. Danielsson, Esko Keski-Vakkuri and Mart\'{\i}n Kruczenski \\
         Institutionen f\"{o}r teoretisk fysik \\
         Box 803\\
         S-751 08  Uppsala \\
         Sweden  \\
         \email{ulf@teorfys.uu.se} \\
         \email{esko@teorfys.uu.se}\\
         \email{martink@teorfys.uu.se}}
         
\preprint{hep-th/9812007,  UUITP-10/98}

\abstract{
In this paper we investigate the relation between the 
bulk and boundary in AdS/CFT. We first discuss the relation
between the Poincar\'e and the global vacua, and then 
study various probes of the bulk from the boundary
theory point of view. 
We derive expressions for retarded propagators and
note that objects in free fall look like expanding bubbles in the
boundary theory.
We also study several Yang-Mills theory examples where we investigate
thermal screening and confinement using propagators.
In the case of confinement we also calculate the profile of
a flux tube and provide an alternative derivation of the tension.
}


\begin{document}

\section{Introduction}

After the discovery of the AdS/CFT correspondence \cite{Maldacena:1997re}
(see also \cite{Polyakov:1998ju,Polyakov:1997tj}) and its more elaborated
formulation in \cite{Gubser:1998bc}\cite{Witten:1998qj}, 
%has been studied in
%numerous subsequent papers. Straightforward 
numerous supergravity calculations have been used to learn more about
supersymmetric Yang-Mills theories. Particularly interesting is the
calculation of Wilson loops in \cite{Rey:1998ik,Maldacena:1998im} which have also been
generalized to finite temperature \cite
{Brandhuber:1998er,Rey:1998bq,Minahan:1998xb,Brandhuber:1998bs,Danielsson:1998br}%
. Ultimately, one hopes to be able to use the correspondence to study also
non supersymmetric Yang-Mills theories, preliminary steps in this direction
have been taken in \cite{Witten:1998zw,Gross:1998gk}. In particular there
exist calculations of glueball masses that agree reasonably well with
lattice calculations \cite{Ooguri:1998hq,deMelloKoch:1998qs,Csaki:1998qr,Hashimoto:1998if}. It
is however far from clear how to connect these models with real QCD in a
rigorous way \cite{Russo:1998mm,Csaki:1998cb}.

In \cite{Balasub:1998sn,Balasub:1998de} the relation between scale in the
boundary and radial position in the bulk was studied in some detail. This is
especially interesting from the point of view of holography, i.e. the
capability of the boundary theory to fully encode the bulk theory. (Note
that this can be expected only if the bulk theory includes gravity, while
the use of a bulk theory to describe a boundary theory can be expected to be
more general). In this paper we will continue the investigation of this
correspondence.

In the first section we clarify some issues concerning vacua defined using
various coordinate systems. Next we derive exact retarded propagators that
can be used to calculate boundary fields for time dependent bulk
configurations. In section three we proceed by considering various
Yang-Mills theory examples. We derive the screening length and the $%
F_{\mu\nu}^{2}$ expectation value at finite temperature using WKB
techniques. For the confining case we calculate the shape of a flux tube and
verify that the integrated energy density gives a tension that reproduces
the confining potential calculated using other means.

\section{The Global and the Poincar\'e Vacuum}

A basic starting point in field theory is the vacuum state. In curved space
such as the AdS space the appropriate choice is often non-trivial and has a
great impact on the physical picture. For example, in investigating the
thermodynamics of black holes it is crucial to impose a correct initial
vacuum state to see the Hawking radiation. An often used vacuum choice in
the AdS/CFT literature is the vacuum corresponding to the Poincar\'e
coordinate system, the Poincar\'e vacuum. For example, the Poincar\'e vacuum
is the chosen initial vacuum state in studies of thermodynamics (see \emph{%
e.g.} \cite
{Maldacena:1998bw,Keski-Vakkuri:1998nw,Muller-Kirsten:1998mt,Ohta:1998xh})
of BTZ black holes \cite{Banados:1992wn,Banados:1993gq}. A slightly puzzling
issue is that the Poincar\'e coordinates cover only a part of the AdS
manifold so there is a coordinate horizon. It would seem preferable to use
coordinates that cover the whole manifold, the global coordinates, and use
the associated global vacuum state as a starting point. It is therefore
important to understand if there is a difference in the two vacuum choices,
and what the difference will be.

In this section we study the relation between the global and the Poincar\'e
vacuum. First we define the vacua in the framework of quantized free bulk
fields and the corresponding operators in the boundary theory, and make a
comparison between the states in the Poincar\'e and global coordinates. We
then move into a more general level and study the Virasoro generators in the
Poincar\'e and global coordinates, and relate the two vacua into each other
using the transformation properties of the Virasoro generators. By the
bulk/boundary correspondence, it should not matter whether the states are
compared using their bulk theory or their boundary theory representation,
since both represent the same Hilbert space. We will therefore mostly focus
on the representation of the states in the boundary theory.

\subsection{AdS coordinates}

In this section we collect some results concerning various coordinate
systems for $AdS$. We will focus on $AdS_{3}$, but the extension of the
results to the general case should be obvious.

$AdS_{3}$ can be defined as the submanifold of $R^{2,2}$ defined by the
equation 
\begin{equation}
U^{2}+V^{2}-X^{2}-Y^{2}=1
\end{equation}
where $U$,$V$,$X$ and $Y$ parameterize the embedding $R^{2,2}$.

In the following we will use three coordinate systems, namely global $(\tau
,\mu ,\theta )$, Poincar\'{e} $(t,x_{0},x)$ and BTZ $(\tilde{t},\phi ,r)$
defined through 
\begin{eqnarray}
U &=&\cosh \mu \cos \tau =\frac{1}{2x_{0}}(1+x^{2}+x_{0}^{2}-t^{2})=r\cosh
(r_{+}\phi -r_{-}t)  \nonumber \\
V &=&\cosh \mu \sin \tau =\frac{t}{x_{0}}=\sqrt{r^{2}-1}\sinh (-r_{-}\phi
+r_{+}t)  \nonumber \\
X &=&\sinh \mu \sin \theta =\frac{x}{x_{0}}=r\sinh (r_{+}\phi -r_{-}t) 
\nonumber \\
Y &=&\sinh \mu \cos \theta =\frac{1}{2x_{0}}(1-x^{2}-x_{0}^{2}+t^{2})=\sqrt{%
r^{2}-1}\cosh (-r_{-}\phi +r_{+}t)
\end{eqnarray}
The BTZ coordinates parameterize only a patch within the Poincar\'{e} space
which is also only part of the whole $AdS_{3}$ space spanned by the global
coordinates. If the identification $\phi \equiv \phi +2\pi $ is made in BTZ
coordinates then the space becomes a black hole and the parameters $r_{+}$
and $r_{-}$ represent the radii of the outer and inner horizons.

The boundary of $AdS$ space is at $\mu \rightarrow \infty $, $%
x_{0}\rightarrow 0$, $r\rightarrow \infty $ in the different coordinates.
Near the boundary the coordinate transformations look like conformal
transformations between boundary variables. It is easy to see that the
conformal transformations are 
\begin{eqnarray}
z &=&\tan \left( \frac{w}{2}\right) =\tanh \left( (r_{+}-r_{-})\frac{\phi
_{+}}{2}\right)  \nonumber \\
\bar{z} &=&\tan \left( \frac{\bar{w}}{2}\right) =\tanh \left( (r_{+}+r_{-})%
\frac{\phi _{-}}{2}\right)
\end{eqnarray}
with 
\begin{equation}
\begin{array}{lclcl}
z=t+x & , & w=\tau +\theta & , & \phi_{+}=\tilde{t} +\phi \\ 
\bar{z}=t-x & , & \bar{w}=\tau -\theta & , & \phi _{-}=\tilde{t} -\phi
\end{array}
\end{equation}

\subsection{Quantization of the Bulk Fields and Boundary Operators}

We begin with a short review of the operator quantization scheme in the bulk
and boundary theory in Minkowski signature, following the references \cite
{Balasub:1998sn}\cite{Balasub:1998de}. A free bulk field is promoted to a
quantized operator through an expansion 
\begin{equation}
\Phi =\sum_{k}a_{k}\phi _{k}(r,\vec{x})\ +\ \mathrm{h.c.}
\end{equation}
where $\phi _{k}$ are the normalizable modes in the bulk, in the chosen
coordinate system, and $a_{k}$ are the associated annihilation operators.
Here $r$ denotes the radial coordinate, and $\vec{x}$ labels the boundary
coordinates.

The vacuum $| 0 \rangle$ is defined as the state that satisfies 
\begin{equation}
a_k | 0 \rangle = 0
\end{equation}
for all annihilation operators. If we use the Poincar\'e coordinates, we
call the resulting vacuum the Poincar\'e vacuum, similarly the global
coordinates define the global vacuum.

The boundary operator $\mathcal{O}$ coupling to the bulk field $\Phi $ has a
similar operator expansion 
\begin{equation}
\mathcal{O=}\sum_{k}b_{k}\tilde{\phi}_{k}(\vec{x})\ +\ \mathrm{h.c.}
\end{equation}
where $\tilde{\phi}_{k}$ are the boundary data extracted from the asymptotic
behavior of the normalizable bulk modes near the boundary: 
\begin{equation}
\phi _{k}\sim r^{2h_{+}}\tilde{\phi}_{k}(\vec{x})\ .
\end{equation}
By the bulk/boundary correspondence, the vacuum defined by 
\begin{equation}
b_{k}|0\rangle =0
\end{equation}
corresponds to the same state in the abstract Hilbert space of the theory as
the vacuum defined using the bulk annihilation operators $a_{k}$. We have
only changed the representation of the Hilbert space from bulk to boundary
theory language.

Let us now take a closer look into the boundary theory in the two coordinate
systems. The boundary is the boundary of the covering space of AdS$_3$ (in
this section we will focus on 2+1 dimensions). It is an infinite cylinder
and can be covered by the global coordinates $\tau \in R,\theta \in [0,2\pi]$
with the periodic identification $\theta \sim \theta + 2\pi$.

In global coordinates the mode expansion has a discrete spectrum, 
\begin{eqnarray}
\mathcal{O} &=& \sum_{n=0}^{\infty }\sum_{l=-\infty }^{\infty }b_{n,l}^{(g)}
e^{-i\omega_{n,l}\tau +il\theta }\ +\ \mathrm{h.c.}  \nonumber \\
&=& \sum_{n=0}^{\infty }\sum_{l=-\infty }^{\infty }b_{h_{n,l},\bar{h}%
_{n,l}}^{(g)}\ e^{-ih_{n,l}w-i\bar{h}_{n,l}\bar{w}}\ +\ \mathrm{h.c.}
\end{eqnarray}
where we used the null coordinates $w=\tau +\theta $, $\bar{w}=\tau -\theta$%
. The weights $h_{n,l},\bar{h}_{n,l}$ are related to the mode frequencies $%
\omega _{n,l}$ and the angular momenta $l$ by 
\begin{eqnarray}
h_{n,l} &=&\frac{1}{2}(\omega _{n,l}-l)  \nonumber \\
\bar{h}_{n,l} &=&\frac{1}{2}(\omega _{n,l}+l)
\end{eqnarray}
and the frequencies $\omega _{n,l}$ are given by 
\begin{equation}
\omega _{n,l}=1+\nu +|l|+2n
\end{equation}
and $\nu =\sqrt{1+m^{2}\Lambda ^{2}}$. 

The global vacuum is given by 
\begin{equation}
b^{(g)}_{n,l}|0\rangle _{g}=0
\end{equation}
and the ''one particle states'' are given by 
\begin{equation}
|n,l\rangle =b^{\dagger (g)}_{n,l}|0\rangle _{g}
\end{equation}
The latter transform according to the (elliptic) representation of SL(2,R)
in the global coordinates \cite
{Natsuume:1996ij,Satoh:1997xf,Maldacena:1998bw,Balasub:1998sn,Vil}.

The Poincar\'e coordinates cover only a finite diamond shaped patch $w\in [
-\pi ,\pi ]$, $\bar{w}\in [ -\pi ,\pi ]$ of the infinite cylindrical
boundary of CAdS$_{3}$. The mode spectrum is continuous, and the operator
expansion takes the form 
\begin{eqnarray}
\mathcal{O} &=& \int^{\infty}_0 d\omega \int^{\infty}_{-\infty} dk \
b_{\omega ,k}^{(P)}e^{-i\omega t +ikx}\ +\ \mathrm{h.c.}  \nonumber \\
&=& \int dh\int d\bar{h}\ b_{h,\bar{h}}^{(P)}e^{-ihz-i\bar{h}\bar{z}}\ +\ 
\mathrm{h.c.}
\end{eqnarray}
in the null coordinates $z=t+x,\bar{z}=t-x$. The weights $(h,\bar{h})$ are
related to the energy $\omega $ and the momentum $k$ by 
\begin{eqnarray}
h &=&\frac{1}{2}(\omega -k)  \nonumber \\
\bar{h} &=&\frac{1}{2}(\omega +k)
\end{eqnarray}
The Poincar\'e vacuum is given by 
\begin{equation}
b_{\omega ,k}^{(P)}|0\rangle _{P}=0
\end{equation}
and the ''one particle states'' are given by 
\begin{equation}
|\omega ,k\rangle =b_{\omega ,k}^{\dagger (P)}|0\rangle _{p}\ .
\end{equation}
They transform according to the (parabolic) representation of SL(2,R) in the
Poincar\'e coordinates \cite
{Natsuume:1996ij,Satoh:1997xf,Maldacena:1998bw,Balasub:1998sn,Vil}.

\subsection{The Relation of the Vacua}

One might worry that since the global vacuum is defined over the whole
infinite cylinder, a transformation to a single finite Poincar\'e patch will
create a mixed state. 

However, it turns out that if the parameter $\nu $ is an integer, the
situation is simpler. One can check that in fact in this case the
operator $\mathcal{O}$ becomes periodic or antiperiodic over a fundamental
domain which is exactly the Poincar\'e diamond with periodic identifications 
$w\sim w+2\pi ,\bar{w}\sim \bar{w}+2\pi $. In other words, we can view the
operator $\mathcal{O}$ as being defined on a torus. 
%$(w,\bar{w})\in \lbrack -\pi ,\pi ]\times \lbrack -\pi ,\pi ]$. 
If the parameter $\nu $ is an \emph{odd} integer, the weights $h,\bar{h}$
are integer valued, and the operator $\mathcal{O}$ is periodic over the
torus. If $\nu $ is an \emph{even} integer, the weights $h,\bar{h}$ are
half-integer valued and the operator $\mathcal{O}$ is antiperiodic over the
torus.

The Poincar\'{e} patch does not quite cover a torus (one would need more
than one patch). To manage with a single patch, we need to cut the torus
open with infinitesimal cuts. If we begin with a global vacuum on the torus,
the cuts in principle will induce a logarithmically divergent entropy due to
the correlations lost across the cut \cite
{Bombelli:1986rw,Srednicki:1993im,Hol}. However, this contribution can be
ignored. Thus, after the cuts the system is still in a global vacuum state.

Now we can simply study how the modes transform under the coordinate
transformation 
\begin{equation}
z=\Lambda \tan (w/2)\ ,\ \bar{z}=\Lambda \tan (\bar{w}/2) 
\end{equation}
\mbox{}from the global coordinates to the Poincar\'e coordinates\footnote{%
Incidentally, the same transformation is used to map the infinite Minkowski
space to a finite Penrose diagram in 1+1 dimensions (see \cite
{Birrell:1982ix}). Thus the global coordinates are analogous to the Penrose
coordinates and the Poincar\'e coordinates are analogous to the Minkowski
coordinates.}. One can easily check that the Poincar\'e modes are analytic
and bounded in the lower complex global time half-plane, so they correspond
to positive energy modes in global coordinates \footnote{%
We thank Jorma Louko for emphasizing this to us.}.

As an aside, we also note that 
the transformation from global to Poincar\'e coordinates
is different from the transformation from Minkowski to Rindler coordinates
in Minkowski space. The $\tau=0$ global time slice coincides with the $t=0$
Poincar\'e time slice, whereas a $t=0$ Minkowski time slice is divided
between two Rindler wedges\footnote{We thank I. Bengtsson and P. Kraus
for pointing this out to us.}.

\bigskip

There is one subtlety which arises from mapping the finite region in global
coordinates to an infinite region in Poincar\'e coordinates. The vacuum will
develop a finite energy density.
Subsequently the whole energy spectrum is shifted by the vacuum energy. We
discuss this in more detail below.

\subsection{The Virasoro Generators}

The discussion in the previous section was based on treating the operators
as free fields. We will now study the vacua again in a more general
framework and examine how Virasoro generators (corresponding to the 
generators of
asymptotic isometries of AdS$_3$ \cite{Brown:1986nw})
are related in the two coordinate systems.

The Virasoro generators are a special case of conserved charges $L_f$
corresponding to currents constructed from the energy momentum tensor: 
\begin{equation}
L_f = \int dz \ T(z) f(z)  \label{charges}
\end{equation}
where the integration is taken along an appropriate contour and $f(z)$ is a
holomorphic function (see \emph{e.g.} paper IV.2 in \cite{Jackiw:1995be} for
more discussion\footnote{%
In what follows, we will be somewhat cavalier and not address issues of
convergence in the definition of the charges.}). The charges satisfy the
commutation relations 
\begin{equation}
i[L_f,L_g] = L_h - \frac{a}{2} \int dz \ f^{\prime\prime\prime}(z) g(z)
\label{commutator}
\end{equation}
where $h=f^{\prime}g-g^{\prime}f$ and the last term is an additional $c$%
-number. In the radial quantization, one takes the integration contour to be
a circle of fixed radius around the origin, then the choice $%
f=-iz^{1-n},g=-iz^{1-m}$ reproduces the Virasoro generators and the
commutation relations of the Virasoro algebra with the central extension
term where $a$ is related to the central charge $c$.

In our case, for the global coordinates on the cylindrical boundary we
choose to define the Virasoro generators as follows: 
\begin{equation}
L^{(g)}_n = \int^{\infty}_{-\infty} dw \ e^{inw} T(w) \ ,
\end{equation}
similarly for the other null coordinate $\bar{w}$ we define $\bar{L}^{(g)}_n$%
. Recall that the mode spectrum is discrete in the global coordinates.

In the Poincar\'e coordinates, the spectrum is continuous, so we define a
continuum of charges $L^{(P)}_{\lambda}$ (similarly $\bar{L}^{(P)}_{\lambda}$%
) by 
\begin{equation}
L^{(P)}_{\lambda} = \int_{-\infty }^{\infty } dz \ e^{i\lambda z}T(z).
\end{equation}
Note that $L^{(P)}_{\lambda =0} +\bar{L}^{(P)}_{\lambda =0}$ is the
Hamiltonian in the Poincar\'e coordinates\footnote{%
Compare this with the radial quantization on the Poincar\'e plane, where the
Hamiltonian is $L_{-1}+\bar{L}_{-1}$, see \emph{e.g.} \cite{Balasub:1998sn}.}%
. Then, the commutation relations (\ref{commutator}) imply that the $%
L^{(P)}_{\lambda \neq 0}$ charges act as raising and lowering operators on
energy eigenstates: 
\begin{equation}
[L^{(P)}_0,L^{(P)}_{\lambda}] = -\lambda L^{(P)}_{\lambda} \ .
\label{ladder}
\end{equation}

Under the coordinate transformation from the global to Poincar\'e
coordinates $w\rightarrow z(w)$, the stress tensor transforms as a dimension
2 operator 
\begin{equation}
T^{P}(z)=\left( \frac{dw}{dz}\right) ^{2}T^{g}(w(z))+\frac{c}{12}S(w,z).
\end{equation}
The first term in the RHS is the classical transformation, which at the
quantum level is supplemented by the second term which arises from the
conformal anomaly. $S(w,z)$ denotes the Schwarzian derivative.

Next, we work out the Schwarzian and the transformation rules for the
individual charges $L$ which we defined above. The results for the latter
turn out to be: 
\begin{equation}
L^{(P)}_{\lambda =0} =L_{n=0}^{(g)}+\frac{1}{2}%
(L_{n=-1}^{(g)}+L_{n=1}^{(g)}) -\frac{c}{24}
\end{equation}
\begin{equation}
L_{\lambda >0}^{(P)}=\sum_{n\geq -1}c_{n}(\lambda )L_{n}^{(g)}
\end{equation}
\begin{equation}
L_{\lambda <0}^{(P)}=\sum_{n\leq 1}\tilde{c}_{n}(\lambda )L_{n}^{(g)}
\end{equation}
where $c_{n}(\lambda ),\tilde{c}_{n}(\lambda )$ are certain coefficients
which can be expressed in terms of hypergeometric functions of $\lambda$.
Note that the first relation (without the $c$-number term) is equal to the
relation between the time translation generating vector field $i\partial_z
+i\partial_{\bar{z}}$ and the vector field representation of the SL(2,R)
generators in the global coordinates.

Acting on the SL(2,R) invariant global vacuum $|0\rangle _{g}$ with the $%
L_{\lambda}^{(P)}$, we obtain the following results: 
\begin{equation}
L_{\lambda =0}^{(P)}|0\rangle _{g}=-\frac{c}{24}
\end{equation}
\begin{equation}
L_{\lambda >0}^{(P)}|0\rangle _{g}=0
\end{equation}
\begin{equation}
L_{\lambda <0}^{(P)}|0\rangle _{g}=\sum_{n\leq -2}\tilde{c}_{n} (\lambda
)L_{n}^{g}|0\rangle _{g}\neq 0.
\end{equation}
The second equation above shows that $|0\rangle _{g}$ is the lowest energy
state, because it is annihilated by all the lowering operators $%
L^{(P)}_{\lambda >0}$. The first equation tells that in the Poincar\'{e}
coordinates the global vacuum state is seen to have a negative energy
density. This can be understood as a shift of the vacuum
energy density associated with
the stretching of the finite area of the patch to an infinite area in the
Poincar\'{e} coordinates. Consequently, the whole energy spectrum is shifted
by the constant vacuum energy. This is a bit reminiscent of the Casimir
effect, however in that case the restriction to the finite domain is obtained
by imposing Dirichlet boundary conditions at the plates and the Casimir
energy density depends on their separation.
The last equation shows that $L_{\lambda
<0}^{P}$ acts as a creation operator on the global vacuum.

\mbox{}From the bulk point of view we can formulate the result in the
following way. It is easy to see that an observer at rest in Poincar\'{e}
coordinates is accelerating with respect to an inertial observer using the
global vacuum. This leads to the naive expectation that the Poincar\'{e}
observer should see a thermal background, but as we have argued this is not
the case. The background is not thermal but instead given by a shifted 
vacuum energy density. 
The acceleration of a Poincar\'{e} observer represents, however, a
critical value such that any larger acceleration will result in a
temperature as is the case for the vacuum defined by BTZ coordinates. See 
\cite{Deser:1998xb} for related results.

To summarize, the Poincar\'e vacuum state can be identified with the global
vacuum state after the negative vacuum energy has been subtracted
off. Hence in investigating quantum fields in AdS space, it does not make
much of a difference whether one starts with a global or a Poincar\'e vacuum
state. In particular, this result gives a more solid footing to the
investigations of AdS black hole thermodynamics \cite
{Maldacena:1998bw,Keski-Vakkuri:1998nw,Muller-Kirsten:1998mt,Ohta:1998xh}.

\section{Bulk to bulk propagators in $AdS$ space}

We now turn our attention to various propagators in the $AdS$ space, these
will be used later for the boundary theory description of different bulk
probes.

The propagator for a scalar field is defined as a solution to the equation 
\begin{equation}
\frac{1}{\sqrt{g}}\partial_{\mu} \left(\sqrt{g} g^{\mu\nu} \partial_{\nu}
G(x,x^{\prime}) \right) - m^2 G(x,x^{\prime}) = \frac{1}{\sqrt{g}}
\delta(x-x^{\prime}) ,  \label{propeqn}
\end{equation}
where $g^{\mu\nu}$ is the metric of the space and $m$ the mass of the field.
Also, $x^{\prime}$ is the position of the source and $x$ the point where the
field is measured. In the case of $AdS_{d+1}$ space with Euclidean signature
the metric is in Poincar\'e coordinates 
\begin{equation}
ds^2 = \frac{1}{x_0^2}\left( dx_0^2 + dx_1^2 +\ldots dx_d^2 \right). 
\end{equation}
In $AdS$, $m^2$ can be negative. A more useful parameter is given by: 
\begin{equation}
\nu = \sqrt{\frac{d^2}{4} + m^2} 
\end{equation}
which must be real to ensure stability, this imposes a lower bound $m^2 \geq
-d^2/4$ for the mass \cite{Breitenlohner:1982jf,Breitenlohner:1982bm,Mezincescu:1985ev}. 
The propagator is a function of the invariant distance $d(x,x^{\prime})$ in 
$AdS$ space, which in the Poincar\'e coordinates is defined by 
\begin{equation}
\cosh(d(x,x^{\prime})) = 1+\frac{(x_0-x^{\prime}_0)^2+\ldots+(x_d-x^{%
\prime}_d)^2}{2x_0x^{\prime}_0} 
\end{equation}
The solution of (\ref{propeqn}) can be found in \cite{Berenstein:1998ij} but
we include it for completeness\footnote{%
This expressions are equivalent to the one in \cite{Berenstein:1998ij} which
uses an hypergeometric function.}: 
\begin{equation}
\begin{array}{ll}
d = 2m & G^{\nu}_{2m}(u) = -\frac{1}{4\pi} \left(-\frac{1}{2\pi\sinh(u)}%
\frac{d}{du}\right)^{m-1} \frac{e^{-u\nu}}{\sinh(u)} \\ 
&  \\ 
d = 2m+1 & G^{\nu}_{2m+1}(u) = -\frac{1}{2\pi}\left(-\frac{1}{2\pi\sinh(u)}%
\frac{d}{du}\right)^{m} Q_{\nu-\frac{1}{2}}\left(\cosh(u)\right)
\end{array}
\label{eq:euprop}
\end{equation}
where $u=d(x,x^{\prime})$ and $Q$ is a Legendre function of the second kind 
\cite{Grad}. These expressions are valid in any coordinate system provided
we substitute the appropriate expression for the invariant distance $u$.

In the Minkowski case there are different propagators depending on the time
boundary conditions, the most relevant ones are the Feynman and the retarded
propagators. The Feynman propagator is the analytic continuation of the
Euclidean one so it can be obtained from (\ref{eq:euprop}). The retarded
propagator can be very different as is illustrated by the well-known example
of $3+1$ massless propagators in flat space. Recall also that in flat space
the massless propagators are concentrated in the light cone in even
dimensions. The Poincar\'e metric in Minkowski signature is given by: 
\begin{equation}
ds^2 = \frac{1}{x_0^2}\left( dx_0^2 -dt^2 + dx_1^2 +\ldots dx_{d-1}^2
\right). 
\end{equation}
The propagator is a function of $x^{\prime}_0$, $x_0$, $t-t^{\prime}$ and $r=%
\sqrt{(x_1-x^{\prime}_1)^2 +\ldots+(x_{d-1}-x^{\prime}_{d-1})^2}$. It
satisfies the equation: 
\begin{equation}
\partial_{00} G + \frac{1-d}{x_0} \partial_0 G -\partial_{tt}G
+\partial_{rr} G + \frac{d-2}{r} \partial_r G +\frac{m^2}{x_0^2} G = \frac{%
\Gamma\left(\frac{d-1}{2}\right)}{2\pi^{\frac{d-1}{2}}} x_0^{d-1} \frac{%
\delta(x_0-x^{\prime}_0) \delta(t-t^{\prime}) \delta(r)}{r^{d-2}}
\label{propeqn2}
\end{equation}
In this equation the dimension $d$ is just a parameter. A simple computation
reveals a relation for propagators in different dimensions but for a fixed
value of $\nu$: 
\begin{equation}
G^{\nu}_{d+2} = -\frac{1}{2\pi}\frac{x_0x^{\prime}_0}{r} \frac{d}{dr}
G^{\nu}_d  \label{eq:dimup}
\end{equation}
The only subtlety in deriving this relation is that in the right hand side
of (\ref{propeqn2}) the radial delta functions have different measures in
different dimensions. Therefore, one must use the relation $%
r\delta^{\prime}(r)=-\delta(r)$ for delta functions acting on functions
regular at $r=0$.

\mbox{} From (\ref{eq:dimup}) it follows that it is only necessary to obtain
the propagators in $AdS_{2+1}$ and $AdS_{3+1}$. The retarded propagator
vanishes for $t<t^{\prime}$ and for $t>t^{\prime}$ it must be a linear
combination of the solutions of the homogeneous equation. Among these
solutions, we will only consider the normalizable ones. This amounts to a
choice of a boundary condition in the boundary of $AdS$ space which seems to
be the most natural one given the formulation of the AdS/CFT correspondence
in Lorentzian signature\cite{Balasub:1998sn}. The normalizable modes can be
written in terms of Bessel functions as \cite{Balasub:1998sn}: 
\begin{equation}
\Phi = e^{-i\omega t+i k x} x_0 J_{\nu}(qx_0); \ \ w^2=q^2+k^2. 
\end{equation}
The linear combination is then chosen in such a way that 
\begin{equation}
\left. \partial_t G \right|_{t^{\prime}\rightarrow t^+} - \left. \partial_t
G \right|_{t^{\prime}\rightarrow t^-} = -\frac{\Gamma\left(\frac{d-1}{2}%
\right)}{2\pi^{\frac{d-1}{2}}} x_0^{d-1} \delta(x_0-x^{\prime}_0) \frac{%
\delta(r)}{r^{d-2}} 
\end{equation}
Using the completeness relation of Bessel functions 
\begin{equation}
\int_0^{\infty} dq q J_{\nu}(qx^{\prime}) J_{\nu}(qx) = \frac{%
\delta(x^{\prime}-x)}{x} 
\end{equation}
the propagator for $t>t^{\prime}$ can be written as 
\begin{equation}
G(x_0,x^{\prime}_0,t-t^{\prime},r) =-\frac{x_0x^{\prime}_0}{\pi}
\int_0^{\infty}\int_0^{\infty} dk\, dq\, q \frac{\sin(\sqrt{q^2+k^2}%
(t-t^{\prime}))}{\sqrt{q^2+k^2}} \cos(kr) J_{\nu}(qx^{\prime}_0)
J_{\nu}(qx_0) 
\end{equation}
Performing the $k$ integral we obtain \cite{Grad} 
\begin{equation}
G(x_0,x^{\prime}_0,t,r) = -\Theta(t-r) \frac{x_0x^{\prime}_0}{2}
\int_0^{\infty} dq\, q J_{\nu}(qx^{\prime}_0) J_{\nu}(qx_0)J_0(q\sqrt{t^2-r^2%
}) 
\end{equation}
where $\Theta(x)$ is the Heaviside function ($\Theta(x>0)=1$ and $%
\Theta(x<0)=0$). The last integral can also be computed \cite{Grad} in terms
of associated Legendre functions of the first and second kind ($%
P^{\alpha}_{\beta},Q^{\alpha}_{\beta}$), with the result: 
\begin{equation}
G(x_0,x^{\prime}_0,t,r) = \left\{ 
\begin{array}{lcl}
-\frac{1}{2\sqrt{2\pi}}\frac{1}{\sqrt{\sin(v)}} P^{\frac{1}{2}}_{\nu-\frac{1%
}{2}}(\cos(v)) & \mbox{if} & \left\{ 
\begin{array}{c}
(x_0-x^{\prime}_0)^2+r^2-t^2<0, \\ 
(x_0+x^{\prime}_0)^2+r^2-t^2>0
\end{array}
\right. \\ 
&  &  \\ 
\frac{i}{\sqrt{2}\pi^{\frac{3}{2}}} \frac{\sin(\nu\pi)}{\sqrt{\sinh(u)}} Q^{%
\frac{1}{2}}_{\nu-\frac{1}{2}}(\cosh(u)) & \mbox{if} & \left\{ 
\begin{array}{c}
(x_0-x^{\prime}_0)^2+r^2-t^2<0, \\ 
(x_0+x^{\prime}_0)^2+r^2-t^2<0
\end{array}
\right. \\ 
&  &  \\ 
0 & \mbox{if} & (x_0-x^{\prime}_0)^2+r^2-t^2>0
\end{array}
\right. 
\end{equation}
where we defined 
\begin{equation}
\begin{array}{lcl}
\cos(v) & = & 1-\frac{t^2-r^2-(x_0-x^{\prime}_0)^2}{2x_0x^{\prime}_0} \\ 
\cosh(u) & = & 1+\frac{t^2-r^2-(x_0-x^{\prime}_0)^2}{2x_0x^{\prime}_0}
\end{array}
\label{eq:cosv}
\end{equation}
Note that $v=d(x,x^{\prime})$ is the timelike distance between $x$ and $%
x^{\prime}$ measured in $AdS$ space.

The case when $\nu $ is integer is particularly simple since $\sin (\nu \pi
)=0$. In that case we obtain 
\begin{equation}
G_{d=2}^{\nu \in Z}(x_{0},x_{0}^{\prime },t,r)=\left\{ 
\begin{array}{ll}
-\frac{\cos (\nu v)}{2\pi \sin (v)} & \mbox{if}-1<\cos (v)<1,\ 
\mbox{i.e. $v$ is
real} \\ 
0 & \mbox{otherwise}
\end{array}
\right.  \label{eq:propads3}
\end{equation}
To obtain the propagator in higher dimensions we apply the identity (\ref
{eq:dimup}). Since the propagator depends only on the invariant distance $v$
(as expected) it is better to rewrite that equation as 
\begin{equation}
G_{d+2}^{\nu }(v)=\frac{1}{2\pi \sin (v)}\frac{d}{dv}G_{d}^{\nu }(v) 
\end{equation}
or 
\begin{equation}
G_{d=2m}^{\nu }(v)=\left( \frac{1}{2\pi \sin (v)}\frac{d}{dv}\right)
^{m-1}G_{d=2}^{\nu }(v) 
\end{equation}
with $G_{2}^{\nu }$ given by (\ref{eq:propads3}) for $\nu $ integer. While
taking the derivatives one must be careful since the condition $-1<\cos
(v)<1 $ implies a discontinuity for the function at $\cos (v)=\pm 1$, this
gives rise to delta functions. To see this better we can rewrite the
propagator for $\nu \in Z$ as 
\begin{equation}
G_{d=2m}^{\nu \in Z}(v)=-\frac{1}{2\pi }\left. \left( -\frac{1}{2\pi }\frac{d%
}{d\eta }\right) ^{m-1}K_{\nu }(\eta )\right| _{\eta =\cos (v)} 
\end{equation}
with 
\begin{equation}
K_{n}(\eta )=\Theta (|\eta |-1)\frac{\cos (n\arccos (\eta ))}{\sqrt{1-\eta
^{2}}}=\Theta (|\eta |-1)\frac{T_{n}(\eta )}{\sqrt{1-\eta ^{2}}} 
\end{equation}
where $\Theta $ is the Heaviside function and $T_{n}$ is a Chebyshev
polynomial of order $n$.

In odd dimensional Minkowski space the retarded propagator fills the
interior of the forward light cone even when the field is massless. In $AdS$
there is a further subtlety, as seen in fig.\ref{fig:adsprop} there is a region in
the future of the source where the propagator vanishes. Furthermore, one can
show that the filled lightcone touches the boundary along a hyperbola that
approaches the (unfilled) lightcone in the boundary of even dimensional
Minkowski space as the source approaches the boundary.

\FIGURE[l]{\epsfysize=5.0truecm \epsffile{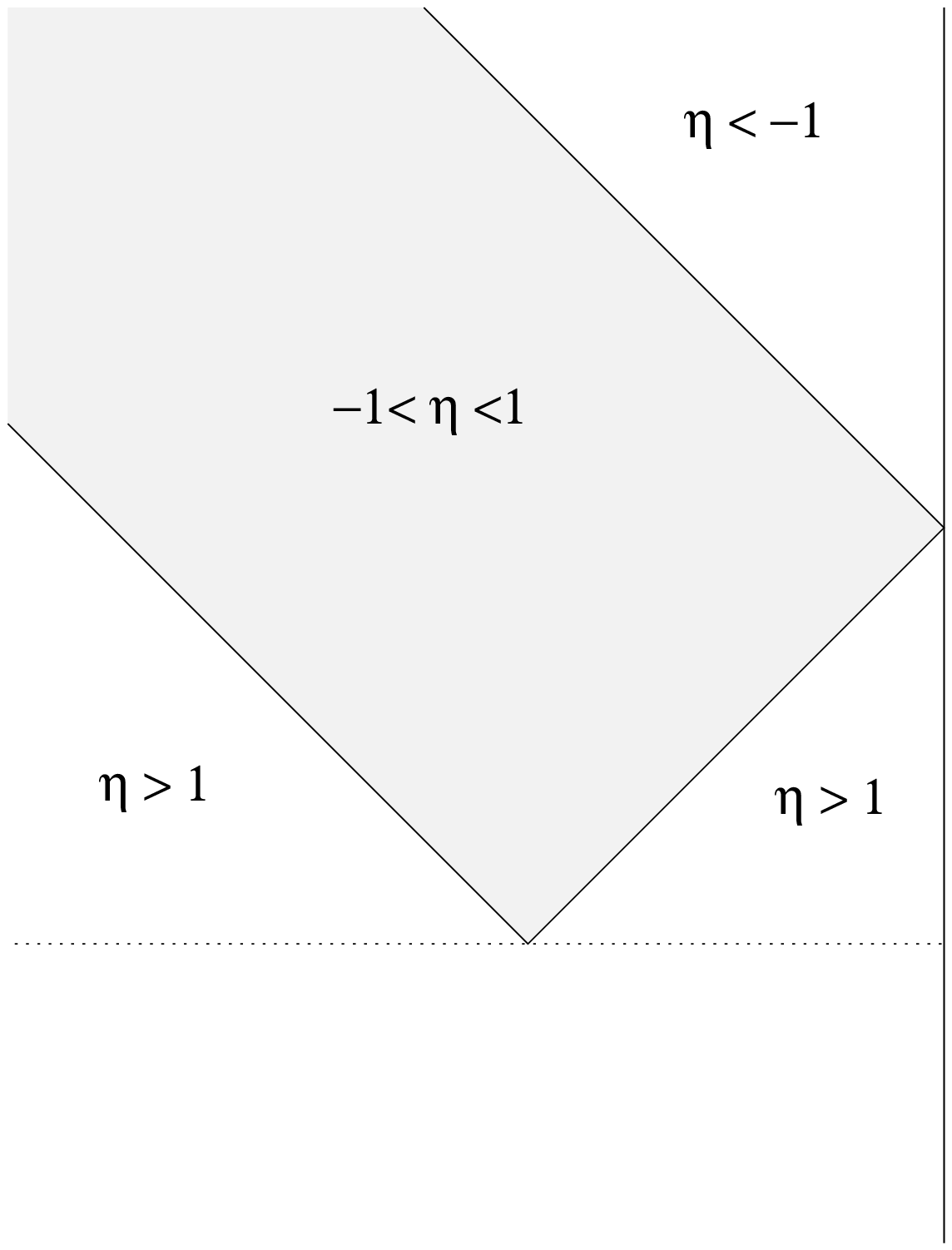}
\caption{For $d=2n+1$ and $\nu\in Z$ the propagator is non-vanishing only in the shaded region.
Here $\eta = \cos v$, see eq.(\ref{eq:cosv}).}\label{fig:adsprop}}

The calculation proceeds in a similar way in the case of $AdS_{3+1}$ \cite{Avis:1978yn}. 
The expansion in modes reads: 
\begin{equation}
G(x_{0},x_{0}^{\prime },t,r)=-\frac{(x_{0}x_{0}^{\prime })^{\frac{3}{2}}}{%
\pi }\int_{0}^{\infty }\int_{0}^{\infty }dk\,dq\,kq\frac{\sin (\sqrt{%
q^{2}+k^{2}}\,t)}{\sqrt{q^{2}+k^{2}}}J_{0}(kr)J_{\nu }(qx_{0}^{\prime
})J_{\nu }(qx_{0}) 
\end{equation}
Performing the $k$ integral we obtain \cite{Grad} 
\begin{equation}
G(x_{0},x_{0}^{\prime },t,r)=-\Theta (t-r)\frac{(x_{0}x_{0}^{\prime })^{%
\frac{3}{2}}}{2\pi }\int_{0}^{\infty }dq\,qJ_{\nu }(qx_{0}^{\prime })J_{\nu
}(qx_{0})\frac{\cos (q\sqrt{t^{2}-r^{2}})}{\sqrt{t^{2}-r^{2}}} 
\end{equation}
This integral is divergent. To handle it we can rewrite it as 
\begin{equation}
\begin{array}{l}
\frac{(x_{0}x_{0}^{\prime })^{\frac{3}{2}}}{2\pi }\left. \frac{d}{\xi d\xi }%
\int_{0}^{\infty }dq\,\sin (q\xi )J_{\nu }(qx_{0}^{\prime })J_{\nu
}(qx_{0})\right| _{\xi =\sqrt{t^{2}-r^{2}}} \\ 
\ \ =\frac{x_{0}x_{0}^{\prime }}{2\pi }\frac{d}{\xi d\xi }\left\{ 
\begin{array}{lcl}
0 & \mbox{if} & \xi <|x_{0}^{\prime }-x_{0}| \\ 
\frac{1}{2}P_{\nu -\frac{1}{2}}\left( \frac{x_{0}^{2}+{x_{0}^{\prime }}%
^{2}-\xi ^{2}}{2x_{0}x_{0}^{\prime }}\right) & \mbox{if} & |x_{0}^{\prime
}-x_{0}|<\xi <x_{0}^{\prime }+x_{0} \\ 
-\frac{\cos (\nu \pi )}{\pi }Q_{\nu -\frac{1}{2}}\left( -\frac{x_{0}^{2}+{%
x_{0}^{\prime }}^{2}-\xi ^{2}}{2x_{0}x_{0}^{\prime }}\right) & \mbox{if} & 
x_{0}^{\prime }+x_{0}<\xi
\end{array}
\right.
\end{array}
\end{equation}
In this case the propagator is simpler for half-integer $\nu $ (which
includes the case $m=0$). Using the notation of eq.(\ref{eq:cosv})
the propagator can be written 
\begin{equation}
G_{d=3}^{\nu \in Z+\frac{1}{2}}(x_{0},x_{0}^{\prime },t,r)=\frac{1}{2\pi }%
\left. \frac{d\tilde{K}_{\nu }(\eta )}{d\eta }\right| _{\eta =\cos (v)} 
\end{equation}
with 
\begin{equation}
\tilde{K}_{\nu }(\eta )=\Theta (|\eta |-1)\frac{1}{2}P_{\nu -\frac{1}{2}%
}(\eta ) 
\end{equation}
Again, using (\ref{eq:dimup}) we can express the propagator in higher
dimensions as: 
\begin{equation}
G_{d=2m+1}^{\nu \in Z+\frac{1}{2}}(v)=-\left. \left( -\frac{1}{2\pi }\frac{d%
}{d\eta }\right) ^{m}\tilde{K}_{\nu }(\eta )\right| _{\eta =\cos (v)} 
\end{equation}
Moreover, in the case we are looking at, namely $\nu -\frac{1}{2}$ integer,
the function $P_{\nu -\frac{1}{2}}$ is a (Legendre) polynomial of order $\nu
-\frac{1}{2}$ and so the derivatives terminate for a given $d$. This means
that the propagator is concentrated on the light-cone, since the only
derivatives that remain are due to the jump of the function.

Finally, all the propagators are independent of a choice of a coordinate
system. Since they transform as scalars, one only needs to express the
invariant distance in the desired coordinates.

\section{Examples}

\subsection{Bubbleography}

Let us now use the propagator which was derived in the last section to study
the boundary theory description of prototypical probes of the bulk geometry.
As a concrete example, we consider a massive source falling along a geodesic
in the AdS$_{d+1}$ space. In global coordinates the metric is given by 
\begin{equation}
ds^{2}=-\cosh ^{2}\mu dt^{2}+d\mu ^{2}+\sinh ^{2}\mu d\Omega _{d-1}^{2}
\end{equation}
and we will consider a source moving along a geodesic at $\mu ^{\prime }=0$.
Primed coordinates will refer to the source, while unprimed coordinates will
refer to the point where we evaluate the field. We are interested in the
leading behavior of a field that couples to the source close to the boundary
where $\mu \rightarrow \infty $ . To find this we need to integrate the
propagator along the worldline of the source, evaluate it in a point a
little bit off the boundary and take the limit $\mu \rightarrow \infty $ and
extract the relevant term. The integral will get contributions from a small
segment of the worldline, $0<v<\pi $ and hence we find 
\begin{equation}
\phi \left( \Omega ,\mu \right) =-\frac{1}{2\pi }\int ds\left( \frac{1}{2\pi
\sin v}\frac{d}{dv}\right) ^{n-1}\frac{\cos \nu v}{\sin v}, 
\end{equation}
where $n=d/2$. We will only consider integer values of $\nu $. For the $\mu
^{\prime }=0$ worldline $\ ds=dt^{\prime }$, and the invariant length is
given by

\begin{equation}
\cos v=\cosh \mu \cos \left( t-t^{\prime }\right) . 
\end{equation}
This implies 
\begin{equation}
\frac{dt^{\prime }}{dv}=-\frac{\sin v}{\cosh \mu }\frac{1}{\sqrt{1-\frac{%
\cos ^{2}v}{\cosh ^{2}\mu }}} 
\end{equation}
and hence 
\begin{equation}
\phi \left( \Omega ,\mu \right) =-\frac{1}{2\pi }\int_{0}^{\pi }dv\frac{\sin
v}{\cosh \mu }\frac{1}{\sqrt{1-\frac{\cos ^{2}v}{\cosh ^{2}\mu }}}\left( 
\frac{1}{2\pi \sin v}\frac{d}{dv}\right) ^{n-1}\frac{\cos \nu v}{\sin v}. 
\end{equation}
The integral over $v$ will pick out powers of $\cosh \mu $ such that the
leading contribution for $\mu \rightarrow \infty $ is 
\begin{equation}
\phi =A_{n,\nu }\left( \cosh \mu \right) ^{-n-\nu },  \label{eq:constfield}
\end{equation}
where $A_{n,\nu }$ is a constant that is zero for $\nu +n-1$ odd. Hence we
conclude that the boundary field is constant.

\bigskip

In the Poincar\'{e} coordinates the situation is different. The worldline is
now given by 
\begin{equation}
x_{0}^{\prime 2}=1+t^{\prime 2}, 
\end{equation}
implying the line element 
\begin{equation}
ds=\frac{dt^{\prime }}{1+t^{\prime 2}}\ . 
\end{equation}
We find 
\begin{equation}
\phi \left( x,x_{0}\right) =-\frac{1}{2\pi }\int_{0}^{\pi }dv\frac{1}{%
1+t^{\prime 2}}\frac{dt^{\prime }}{dv}\left( \frac{1}{2\pi \sin v}\frac{d}{dv%
}\right) ^{n-1}\frac{\cos \nu v}{\sin v}\ . 
\end{equation}
If we use the expression for the worldline in the expression for the
invariant length 
\begin{equation}
\cos v=1+\frac{(x_{0}-x_{0}^{\prime })^{2}+x^{2}-\left( t-t^{\prime }\right)
^{2}}{2x_{0}^{\prime }x_{0}} 
\end{equation}
(we have chosen $x^{\prime }=0$) and define 
\begin{equation}
\sin \alpha =\frac{t^{\prime }}{\sqrt{1+t^{\prime 2}}} 
\end{equation}
\begin{equation}
\sin \beta =\frac{t^{2}-x_{0}^{2}-x^{2}-1}{\sqrt{4t^{2}+\left(
t^{2}-x_{0}^{2}-x^{2}-1\right) ^{2}}} 
\end{equation}
we obtain 
\begin{equation}
\sin \left( \alpha +\beta \right) =\frac{-2x_{0}\cos v}{\sqrt{4t^{2}+\left(
t^{2}-x_{0}^{2}-x^{2}-1\right) ^{2}}}. 
\end{equation}
We can now obtain (noting that $\beta $ is independent of $v$) 
\begin{eqnarray}
\phi \left( x,x_{0}\right) &=&-\frac{1}{2\pi }\int_{0}^{\pi }dv\frac{d\alpha 
}{dv}\left( \frac{1}{2\pi \sin v}\frac{d}{dv}\right) ^{n-1}\frac{\cos \nu v}{%
\sin v} \\
&=&-\int_{0}^{\pi }dvA\sin v\frac{1}{\sqrt{1-A^{2}\cos ^{2}v}}\left( \frac{1%
}{2\pi \sin v}\frac{d}{dv}\right) ^{n-1}\frac{\cos \nu v}{\sin v}
\end{eqnarray}
where 
\begin{equation}
A=\frac{-2x_{0}}{\sqrt{4t^{2}+\left( t^{2}-x_{0}^{2}-x^{2}-1\right) ^{2}}} 
\end{equation}
The $v$ integration results in 
\begin{equation}
\phi =B_{n,\nu }\frac{x_{0}^{n+\nu }}{\left( 4t^{2}+\left(
t^{2}-x_{0}^{2}-x^{2}-1\right) ^{2}\right) ^{\frac{n+\nu }{2}}}, 
\end{equation}
which implies a boundary field expectation value of the form 
\begin{equation}
\left\langle \mathcal{O}\right\rangle =C_{n,\nu }\frac{1}{\left(
4t^{2}+\left( t^{2}-x_{0}^{2}-x^{2}-1\right) ^{2}\right) ^{\frac{n+\nu }{2}}}%
, 
\end{equation}
where $B_{n,\nu }$ and \ $C_{n,\nu }$are constants that are zero for $\nu
+n-1$ odd. As an example of the scale/distance duality between the boundary
and the bulk, in \cite{Balasub:1998de} it was shown that an object at a
fixed distance from a horizon would look like an extended blob in the
boundary theory; the closer to the horizon the larger it will appear. A
freely falling object is then expected to grow as it closes in on the
horizon. The above calculation shows that this is not the whole story. Even
if the boundary description of the object starts out like a blob it will
develop into an expanding bubble over time as illustrated in fig.\ref
{fig:bubbles}. This can be understood intuitively in the following way.
Since the object is falling, all scales will expand but at different rates.
The points in the center of the bubble depend on the source at later times,
when the information has not had time to spread much in the transverse
directions. On the other hand, the source is falling faster at later times
and therefore corresponds to a region in the boundary that must expand
faster. The result is a shockwave that produces a bubble.

In the case of AdS$_{3}$ we can also verify that the change of coordinates
from global to Poincar\'{e} amounts to a conformal transformation in the
boundary; the bubble is hence a conformal transformation of the constant
profile. In fact, 
\begin{equation}
4t^{2}+\left( t^{2}-x^{2}-1\right) ^{2}=(1+z^{2})(1+\overline{z}^{2}) 
\end{equation}
(with $z$ as in the previous section) and since 
\begin{equation}
\frac{\partial w}{\partial z}=\frac{2}{1+z^{2}} 
\end{equation}
we find that the bubble is simply given by the conformal scaling prefactors 
\begin{equation}
\left\langle \mathcal{O}\right\rangle =C_{n,\nu }\left( \frac{1}{4}\frac{%
\partial w}{\partial z}\frac{\partial \overline{w}}{\partial \overline{z}}%
\right) ^{\frac{n+\nu }{2}}. 
\end{equation}

As we see, in the case of $AdS_{3}$, a coordinate transformation in the bulk
induces a conformal transformation in the boundary and the boundary field
transforms accordingly. The geodesic $\mu ^{\prime }=0$ where the particle
is always at the center in global coordinates was particularly simple.
However any other geodesic can be obtained from that one using an isometry
of $AdS_{3}$ space and so the boundary field produced by such particle can be
obtained by the corresponding conformal transformation in the boundary.

In Poincar\'{e} coordinates the trajectory is $x_{0}^{\prime 2}=1+t^{\prime
2}$. Performing the isometry $x_{0}\rightarrow x_{0}/a$, $t\rightarrow t/a$,$%
x\rightarrow x/a$, with $a$ an arbitrary parameter we obtain the new
trajectory defined by 
\begin{equation}
x_{0}^{\prime 2}=a^{2}+t^{\prime 2} 
\end{equation}
In the boundary, the conformal transformation induced by these coordinates
is simply a rescaling of $z,\bar{z}$. However in global coordinates the
conformal transformation is not so trivial and is given by 
\begin{equation}
\tan \left( \frac{w^{\prime }}{2}\right) =a\tan \left( \frac{w}{2}\right) ,\
\ \ \tan \left( \frac{\bar{w}^{\prime }}{2}\right) =a\tan \left( \frac{\bar{w%
}}{2}\right) 
\end{equation}
where $w^{\prime }$ are the new coordinates. Since in the coordinates $w$
the boundary field is a constant we obtain the field produced by the
particle following the new geodesic as 
\begin{eqnarray}
\phi (w^{\prime },\bar{w}^{\prime }) &=&\left( \frac{dw}{dw^{\prime }}%
\right) ^{\frac{\nu +1}{2}}\left( \frac{d\bar{w}}{d\bar{w}^{\prime }}\right)
^{\frac{\nu +1}{2}}\,1 \\
&=&\frac{a^{\nu +1}}{\left( a^{2}\cos ^{2}\frac{w^{\prime }}{2}+\sin ^{2}%
\frac{w^{\prime }}{2}\right) ^{\frac{\nu +1}{2}}\left( a^{2}\cos ^{2}\frac{%
\bar{w}^{\prime }}{2}+\sin ^{2}\frac{\bar{w}^{\prime }}{2}\right) ^{\frac{%
\nu +1}{2}}}
\end{eqnarray}
As a check it can be seen that the same expression is obtained if one plugs
in the new geodesic in the previous formulas with the propagator. The
resulting field looks for small times also like an expanding bubble but is
of very different nature since it is obviously periodic in time.

A more interesting case is that of BTZ coordinates. The conformal
transformation from the coordinates $w$ where the field is constant to BTZ
coordinates reads: 
\begin{equation}
\tanh\left((r_+-r_-)\frac{\phi_+}{2}\right) = a \tan\left(\frac{w}{2}\right)
,\ \ \ \tanh\left((r_++r_-)\frac{\phi_-}{2}\right) = a \tan\left(\frac{\bar{w%
}}{2}\right) 
\end{equation}
and the field is accordingly 
\begin{equation}
\phi(\phi_+,\phi_-) =\frac{a^{\nu+1}(r_+^2-r_-^2)^{\frac{\nu+1}{2}}}{%
\left(a^2+(1+a^2)\sinh^2\frac{(r_+-r_-)\phi_+}{2}\right)^{\frac{\nu+1}{2}}
\left(a^2+(1+a^2)\sinh^2\frac{(r_+-r_-)\phi_-}{2}\right)^{\frac{\nu+1}{2}}} 
\end{equation}
In the black hole case, we must perform the identification $\phi=\phi+2\pi$
which amounts to a sum over images: 
\begin{equation}
\phi_{\mbox{BTZ}}(\phi_+,\phi_-) = \sum_{n=-\infty}^{\infty} \phi(\phi_+
+2\pi n,\phi_-+2\pi n) 
\end{equation}
The field is depicted in fig.\ref{fig:bubbles} where $\phi=-\pi\ldots\pi$.

\FIGURE{
\epsfysize=5.0truecm \epsffile{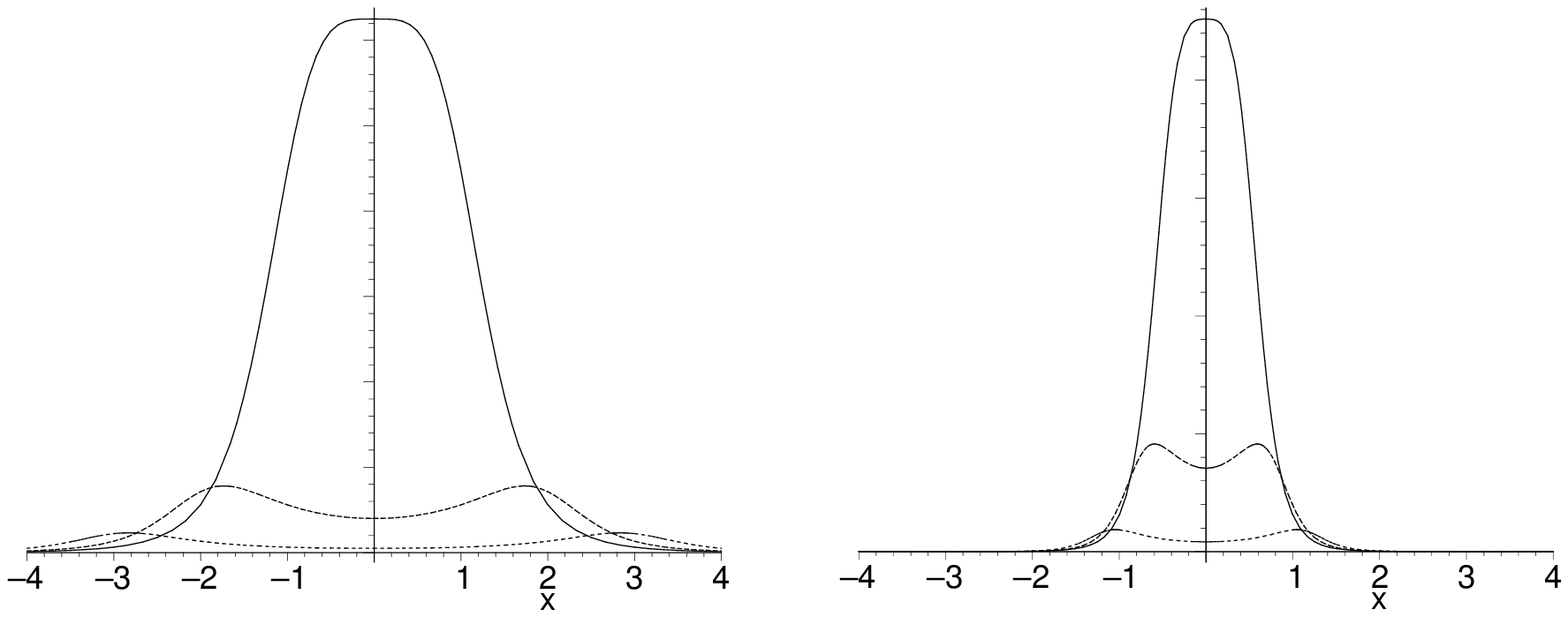}
\caption{Dilaton field in the boundary of, left: $AdS_3$ for $\protect\nu=2$
and times $t=1,2,3$ and right: a BTZ black hole for $\protect\nu=2 $, $r_+=1$%
, $r_-=0$ and times $t=0.5,0.8,1.2$.}
\label{fig:bubbles}}

\subsection{Supersymmetric YM at zero temperature}

\mbox{ }From equation (\ref{eq:constfield}) it follows that the above
picture is no longer relevant if $\nu -n+1$ is odd. In these cases the
expectation value in the boundary theory vanishes. In particular this is
true for massless dilatons, which in the case of $AdS_{5}$ implies that the
expectation value $<F^{2}>$ must vanish for a bulk source in a free fall.
How can one then understand the nonvanishing expectation value obtained in 
\cite{Balasub:1998de}?

The important point is that in \cite{Balasub:1998de} the source is held at a
fixed position in Poincar\'{e} coordinates. This can be achieved by
suspending it on a string hanging down from the boundary with an appropriate
tension. The other endpoint of the string will then show up as a point
charge in the boundary theory. We conclude that nonvanishing expectation
values of $F^{2}$ must be supported by charges. To be more precise we can
use the propagator derived above to calculate $\langle F^{2}\rangle $ due to
a point charge by representing the charge by a string that goes straight
down to the horizon. We must then consider the string as an extended source
for the dilaton field and integrate the propagator along the string.

To perform our calculations we must know how $F^{2}$ couples to the
supergravity fields. There are two such fields that will be of interest to
us, the dilaton $\phi $, and the volume. We will be interested in variations
with respect to certain linear combinations of these fields and a convenient
basis is to consider variations either with the string metric or the
Einstein metric held fixed. To fix the conventions, let us consider the
D-brane action 
\begin{equation}
I_{BI}=Tr\mathcal{T}_{p}\int d^{p+1}xe^{-\phi }\sqrt{\det (g_{\mu \nu }+%
\mathcal{F}_{\mu \nu }+g^{ij}\partial _{\mu }X_{i}\partial _{\nu }X_{j})}. 
\end{equation}
Following e.g. \cite{Chepelev:1997vx,Hashimoto:1998if} we then rescale the worldvolume gauge
field strength, 
\begin{equation}
\mathcal{F}=\mathcal{T}F\ , 
\end{equation}
where $\mathcal{T}$ is the string tension, 
\begin{equation}
\mathcal{T}=\frac{1}{2\pi \alpha ^{\prime }}\ , 
\end{equation}
and substitute the Dp-brane tension 
\begin{equation}
\mathcal{T}_{p}=g_{s}^{-1}(2\pi )^{\frac{1-p}{2}}\mathcal{T}^{\frac{p+1}{2}%
}. 
\end{equation}
Expanding to second order in $F$ we find 
\begin{equation}
I_{BI}=Tr\int d^{p+1}x\sqrt{-g}e^{-\phi }\left( \mathcal{T}_{p}+\frac{1}{%
4g_{YM}^{2}}\left( g^{\mu \nu }F_{\mu \nu }\right) ^{2}+...\right) , 
\end{equation}
where the Yang-Mills coupling is given by 
\begin{equation}
g_{YM}^{2}=(2\pi )^{p-2}g_{s}\alpha ^{\prime }{}^{\frac{p-3}{2}}\ .
\end{equation}
It is convenient to define a new metric $\widehat{g}_{\mu \nu }$, defined
through 
\begin{equation}
g_{\mu \nu }=e^{\frac{2}{p+1}\phi }\widehat{g}_{\mu \nu },
\end{equation}
which for $p=3$ coincides with the Einstein metric. The Born-Infeld action
then becomes

\begin{equation}
I_{BI}=Tr\int d^{p+1}x\sqrt{-\widehat{g}}\left( \mathcal{T}_{p}+\frac{1}{%
4g_{YM}^{2}}e^{^{-\frac{4}{p+1}\phi }}\left( \widehat{g}^{\mu \nu }F_{\mu
\nu }\right) ^{2}+...\right) .
\end{equation}
If we furthermore define 
\begin{equation}
\widehat{\phi }=\frac{4}{p+1}\phi 
\end{equation}
we find that $F^{2}$ is obtained as 
\begin{equation}
\frac{\delta I_{BI}}{\delta \widehat{\phi }}=-\frac{1}{4g_{YM}^{2}}F^{2}.
\label{eq:variation}
\end{equation}
For $p=3$ and $p=4$, we have checked that $\widehat{\phi}$ fluctuations keeping 
$\widehat{g}_{\mu\nu}$ fixed are massless. On the
supergravity side, the supergravity action (dimensionally reduced over the
unit sphere) is then given by 
\begin{equation}
S=-\frac{\Omega _{8-p}}{4\kappa _{10}^{2}}\int d^{p+2}x\sqrt{-g_{E}}%
g_{E}^{\mu \nu }\partial _{\mu }\widehat{\phi }\partial _{\nu }\widehat{\phi 
},
\end{equation}
where $\Omega _{8-p}$ is the volume of the unit $8-p$ sphere. 
If we have a string as a source we will need to vary 
\begin{equation}
S_{int}=\frac{1}{2\pi \alpha ^{\prime }}\int d^{2}x\sqrt{-g}=\frac{1}{2\pi
\alpha ^{\prime }}\int d^{2}xe^{\widehat{\phi }/2}\sqrt{-\widehat{g}},
\label{eq:kalla}
\end{equation}
where the metrics are now the induced ones on the worldsheet. We see that
the variation with respect to $\widehat{\phi }$ gives 
\begin{equation}
\delta S_{int}=\frac{1}{4\pi \alpha ^{\prime }}\int d^{2}x\sqrt{-g}\delta 
\widehat{\phi }.
\end{equation}
The corresponding variation of the supergravity action is then given by the
boundary term 
\begin{equation}
\delta S=\frac{\Omega _{8-p}}{2\kappa _{10}^{2}}\left. \int d^{p+1}x\sqrt{-g}
g^{UU}e^{-2\bar{\phi}}\partial _{U}\widehat{\phi }\delta \widehat{\phi }
\right| _{U\rightarrow \infty }
\end{equation}
where $\bar{\phi}$ is the background value of the dilaton.
Hence it follows that 
\begin{equation}
\frac{1}{4g_{YM}^{2}}\left\langle F^{2}\right\rangle =\left. \frac{\Omega
_{8-p}}{2\kappa _{10}^{2}}\sqrt{-g}g^{UU}e^{-2\bar{\phi}}\partial _{U}%
\widehat{\phi }\right| _{U\rightarrow \infty }  \label{eq:F2dilrel}
\end{equation}
if we consider string like sources.

Let us apply this to the case of a string hanging straight down when $p=3$,
i.e, $d=4$. The string traces out a worldsheet that gives rise to a $\phi $
field 
\begin{equation}
\phi (U,x)=-\frac{1}{4\pi \alpha ^{\prime }}\frac{2\kappa _{10}}{\Omega _{5}}%
\int dt^{\prime }dU^{\prime }\sqrt{-g_{2}}\frac{1}{2\pi \sin v}\frac{d}{dv}%
\frac{\cos 2v}{\sin v} 
\end{equation}
If we want to write the AdS$_{5}$ metric as 
\begin{equation}
ds^{2}=\alpha ^{\prime }{}\left( \left( \frac{R}{U}\right) ^{2}dU^{2}+\left( 
\frac{U}{R}\right) ^{2}\left( \sum_{i=1}^{3}dx_{i}^{2}-dt^{2}\right)
+R^{2}d\Omega _{5}^{2}\right) 
\end{equation}
where $R^{2}=g_{YM}\sqrt{N}$ we need to take $x\rightarrow x/R^{2}$, $%
t\rightarrow t/R^{2}$, rescale the metric by $\alpha ^{\prime }R^{2}$ and
put $x_{0}=1/U$. Noting that $\delta ^{(3)}\left( x\right) =\frac{1}{R^{6}}%
\delta ^{(3)}\left( x/R^{2}\right) $ we find 
\begin{equation}
\phi (U,x)=-\frac{1}{8\pi ^{2}R^{6}}\frac{2\kappa _{10}}{\Omega _{5}}\int
dvdU^{\prime }\frac{U^{-1}U^{\prime -1}}{\sqrt{U^{-2}+U^{\prime -2}+\frac{%
r^{2}}{R^{4}}-2U^{-1}U^{\prime -1}\cos v}}\frac{d}{dv}\frac{\cos 2v}{\sin v} 
\end{equation}
where we have put $r^{2}=\sum_{i=1}^{3}x_{i}^{2}$ and also used 
\begin{equation}
\frac{t-t^{\prime }}{R^{2}}=\sqrt{U^{-2}+U^{\prime -2}+\frac{r^{2}}{R^{4}}%
-2U^{-1}U^{\prime -1}\cos v} 
\end{equation}
We find 
\begin{equation}
\phi (U,x)=-\frac{15}{256\pi ^{2}R^{6}}\frac{2\kappa _{10}}{\Omega _{5}}\int
dU^{\prime }\frac{U^{-4}U^{\prime -4}}{\left( U^{\prime -2}+\frac{r^{2}}{%
R^{4}}\right) ^{7/2}}=-\frac{1}{128\pi ^{2}}R^{2}U^{-4}\frac{1}{r^{4}}\ , 
\end{equation}
hence 
\begin{equation}
\frac{1}{4g_{YM}^{2}}F^{2}=\frac{1}{32\pi ^{2}}g_{YM}\sqrt{N}\frac{1}{r^{4}}
\label{eq:pzerot}
\end{equation}
which suggests that the effective charge goes like $\sqrt{g_{YM}\sqrt{N}}$ . Hence
the force on another charge $\sqrt{g_{YM}\sqrt{N}}$ is $\sim g_{YM}\sqrt{N}\frac{1%
}{r^{2}}$ implying a potential $\sim g_{YM}\sqrt{N}\frac{1}{r}$, in agreement
with \cite{Maldacena:1998im}.

\subsection{Supersymmetric YM at nonzero temperature}

We may also consider the nonzero temperature case. The starting point is the
metric of a non-extremal p-brane, given by 
\begin{eqnarray}
\frac{ds^2}{\alpha^{\prime}}&=&\left( \frac{R}{U}\right)^{\frac{7-p}{2}}%
\frac{dU^{2}}{f(U)}-\left(\frac{U}{R}\right)^{\frac{7-p}{2}} f(U) dt^{2}
+\left(\frac{U}{R}\right)^{\frac{7-p}{2}}\sum_{i=1}^{p}dx_{i}^{2} +R^{\frac{%
7-p}{2}}U^{\frac{p-3}{2}}d\Omega_{8-p}^{2} \nonumber \\
f(U) &=& 1-\frac{U_{T}^{7-p}}{U^{7-p}}  \label{eq:genbhmetric}
\end{eqnarray}
while the dilaton is given by 
\begin{equation}
e^{\phi }=g_{s}\alpha ^{\prime }{}^{\frac{p-3}{2}}\left( \frac{R^{7-p}}{%
U^{7-p}}\right) ^{\frac{3-p}{4}} 
\end{equation}
We have defined 
\begin{equation}
R^{\frac{7-p}{2}}=g_{YM}\sqrt{d_{p}N} 
\end{equation}
where 
\begin{equation}
d_{p}=2^{7-2p}\pi ^{\frac{9-3p}{2}}\Gamma (\frac{7-p}{2}) 
\end{equation}
and $g_{YM}$ is given as before.

We will limit ourselves to the case of $d=4$ and begin by deriving the
propagator for a static configuration since we can no longer use the exact
retarded propagator evaluated in the previous section.

The metric (\ref{eq:genbhmetric}) reduces to 
\begin{eqnarray}
ds^{2} &=&\frac{R^{2}}{U^{2}f(U)}dU^{2}-\frac{U^{2}f(U)}{R^{2}}dt^{2}+\frac{%
U^{2}}{R^{2}}dx_{i}^{2}+R^{2}d\Omega _{5}^{2} \\
f(U) &=&1-\left( \frac{U_{T}}{U}\right) ^{4}
\end{eqnarray}
In this case we will consider only the case of a static source coupled to a
massless field $\phi $ through the action 
\begin{equation}
S_{\mbox{int}}=\int ds\phi (x(t))=\int dt\sqrt{|g_{tt}|}\phi (x(t)) 
\end{equation}
In the propagator equation (\ref{propeqn}), this amounts to multiplying the
right hand side by $\sqrt{|g_{tt}|}$ and omitting the time derivatives. The
resulting equation is 
\begin{equation}
\frac{1}{U^{3}}\partial _{U}\left( \frac{U^{5}f(U)}{R^{2}}\partial
_{U}G(U,U^{\prime },x_{i})\right) +\frac{R^{2}}{U^{2}}\partial
_{ii}G(U,U^{\prime },x_{i})=\frac{\sqrt{f(U)}}{R^{3}U^{2}}\delta
(U-U^{\prime })\delta ^{(3)}(x_{i})  \label{eq:propeqbh}
\end{equation}
The boundary conditions are that $G$ is regular at $U=U_{T}$ and that $G$
vanishes at $U\rightarrow \infty $. 
%since there the metric is regular (in fact looks 
%like flat space since this is an euclidean black hole),
Rescaling the coordinates as $U\rightarrow U_{T}u$, $t\rightarrow
tR^{2}/U_{T}$ and $x\rightarrow xR^{2}/U_{T}$, the equation simplifies to 
\begin{equation}
(u^{5}-u)\partial _{uu}G+(5u^{4}-1)\partial _{u}G+u\partial _{ii}G=\frac{%
\sqrt{u^{4}-1}}{R^{7}u}\delta (u-u^{\prime })\delta ^{(3)}(x_{i}) 
\end{equation}
Fourier transforming both sides of eq.(\ref{eq:propeqbh}) with respect to $%
x_{i}$ leads to 
\begin{equation}
(u^{5}-u)\partial _{uu}\tilde{G}+(5u^{4}-1)\partial _{u}\tilde{G}-uk^{2}%
\tilde{G}=\frac{\sqrt{u^{4}-1}}{R^{7}u}\delta (u-u^{\prime }) 
\end{equation}
with 
\begin{equation}
\tilde{G}(u,u^{\prime },k)=\int d^{3}ke^{i\vec{k}\vec{x}}G(u,u^{\prime
},x_{i}) 
\end{equation}
For $u<u^{\prime }$ and for $u>u^{\prime }$ the Green function is given by
(different) solutions of the homogeneous equation 
\begin{equation}
(u^{5}-u)y^{\prime \prime }(u)+(5u^{4}-1)y^{\prime }(u)-uk^{2}y(u)=0
\label{eq:homeq}
\end{equation}
The homogenous equation has been obtained in \cite{Csaki:1998qr} and
discussed at length in \cite{Zyskin:1998tg}. For $u>u^{\prime }$ there are
two solutions which we shall call $y_{1}(u,k)$ and $\tilde{y}_{1}(u,k)$
which satisfy 
\begin{eqnarray}
y_{1}(u,k) &=&\frac{1}{u^{4}}+\mathcal{O}(\frac{1}{u^{5}}),\ \ u\rightarrow
\infty \\
\tilde{y}_{1}(u,k) &=&1+\mathcal{O}(\frac{1}{u},\frac{\ln (u)}{u^{4}}),\ \
u\rightarrow \infty \\
&&
\end{eqnarray}
and for $u<u^{\prime }$ there is another pair of solutions $y_{2}(u,k)$, $%
\tilde{y}_{2}(u,k)$ satisfying 
\begin{eqnarray}
y_{2}(u,k) &=&1+\mathcal{O}(u-1),\ \ u\rightarrow 1 \\
\tilde{y}_{2}(u,k) &=&\ln (u-1)+\mathcal{O}(u-1,(u-1)\ln (u-1)),\ \
u\rightarrow 1 \\
&&
\end{eqnarray}
The boundary conditions then require that 
\begin{equation}
\tilde{G}(u>u^{\prime },k)=Ay_{1}(u,k),\ \ \tilde{G}(u<u^{\prime
},k)=By_{2}(u,k), 
\end{equation}
and $A$ and $B$ should be chosen such that 
\begin{eqnarray}
By_{2}(u^{\prime },k)-Ay_{1}(u^{\prime },k) &=&0 \\
By_{2}^{\prime }(u^{\prime },k)-Ay_{1}^{\prime }(u^{\prime },k) &=&\frac{1}{%
R^{7}u^{2}\sqrt{u^{4}-1}}
\end{eqnarray}
It follows that 
\begin{eqnarray}
A &=&\frac{y_{1}}{y_{1}y_{2}^{\prime }-y_{2}y_{1}^{\prime }}\frac{1}{%
R^{7}u^{2}\sqrt{u^{4}-1}} \\
B &=&\frac{y_{2}}{y_{1}y_{2}^{\prime }-y_{2}y_{1}^{\prime }}\frac{1}{%
R^{7}u^{2}\sqrt{u^{4}-1}}
\end{eqnarray}
where the functions are evaluated at $u=u^{\prime }$. As follows from (\ref
{eq:homeq}), the Wronskian $W(y_{1},y_{2})=y_{1}y_{2}^{\prime
}-y_{2}y_{1}^{\prime }$ is given by 
\begin{equation}
W(y_{1},y_{2})=\frac{w(k)}{u^{5}-u}  \label{eq:defwk}
\end{equation}
where $w(k)$ is a function that depends only on $k$. This simplifies the
expression for the propagator which turns out to be 
\begin{equation}
\tilde{G}(u,u^{\prime },k)=\frac{1}{w(k)}y_{1}(u_{>},k)y_{2}(u_{<},k)\frac{%
\sqrt{{u^{\prime }}^{4}-1}}{u^{\prime }R^{7}}  \label{eq:propbh}
\end{equation}
with $u_{<}=\mbox{min}(u,u^{\prime })$ and $u_{>}=\mbox{max}(u,u^{\prime })$%
. In ref. \cite{Zyskin:1998tg} a series expansion was obtained for $y_{1}$
and $y_{2}$ and also for $w(k)$. There it was shown that for real $k$ there
are no zeros of $w$ (and so the propagator is well-defined) and also that
there is an infinite number of zeros on the imaginary axes. These zeros give
precisely the masses of the glueballs in \cite{Csaki:1998qr}.

For what follows, besides the expansion of \cite{Zyskin:1998tg} it will be
useful to have also an approximate but simpler expression for $w(k)$ which
we derive in the Appendix using the WKB method.

A charge in the boundary is represented in the bulk by a string hanging from
the boundary down to the horizon. The minimal action configuration
corresponds to the string extended only in the $U$ and $t$ directions. Such
a string will be a source for the dilaton through the coupling 
\begin{equation}
S_{\mbox{int}}=\frac{1}{4\pi }\int dUdt\sqrt{g_{tt}g_{UU}}\phi =\frac{1}{%
4\pi }\int dudt\sqrt{g_{tt}}\frac{uR}{\sqrt{u^{4}-1}}\phi 
\end{equation}
where we have put $\alpha ^{\prime }=1$ for simplicity. The boundary value
of $\phi $ corresponds to the expectation value of $F^{2}$ in the presence
of the charge. We know that for short distances it will behave as $1/r^{4}$
since in that region we can replace the black hole backround by $AdS$ space
and so (\ref{eq:pzerot}) is reproduced. For large distances it should decay
exponentially as $\exp (-r/L_{S})$ where $L_{S}$ is the screening length.
The value of the dilaton can be computed using the static Green function as 
\begin{equation}
\phi (u,x_{i})=\frac{1}{4\pi }\int du^{\prime }\int \frac{d^{3}k}{(2\pi )^{3}%
}e^{-i\vec{k}\vec{x}}\frac{uR}{\sqrt{u^{4}-1}}\tilde{G}(u,u^{\prime },k) 
\end{equation}
Using (\ref{eq:propbh}) the value near the boundary is given by 
\begin{eqnarray}
\phi (U,x)\approx _{U\rightarrow \infty } &&\frac{1}{U^{4}}\phi _{0}(x) \\
\phi _{0}(x) &=&\frac{U_{T}^{4}}{4\pi R^{6}}\int du^{\prime }\int \frac{%
d^{3}k}{(2\pi )^{3}}e^{-i\vec{k}\vec{x}}\frac{y_{2}(u^{\prime },k)}{w(k)}
\end{eqnarray}
The angular integral in the Fourier transform can be performed with the
result 
\begin{equation}
\phi _{0}(x)=\frac{U_{T}^{4}}{(2\pi )^{3}R^{6}}\int du^{\prime
}\int_{0}^{\infty }dk\frac{\sin (kr)}{kr}\frac{y_{2}(u^{\prime },k)}{w(k)} 
\end{equation}
where $r^{2}+x_{1}^{2}+x_{2}^{2}+x_{3}^{2}$. The integral in $k$ can be
extended from $-\infty \rightarrow \infty $ and evaluated by the method of
residues. The integrand has simple poles in the zeros of $w(k)$ which are
located on the imaginary axis. A zero in $w(k)$ means that the solutions $%
y_{1}$ and $y_{2}$ are proportional, that is that for that value of $k$
there exists a solution well behaved in $u=1$ and $u\rightarrow \infty $.
Such solutions are the eigenfunctions computed in \cite
{Csaki:1998qr,Zyskin:1998tg} and so, the zeros of $w(m_{n})=0$ define the
glueball masses in the $2+1$ confining theory. 
\begin{equation}
\phi _{0}(x)=\frac{U_{T}^{4}}{8\pi R^{6}}\sum_{n=1}^{\infty }\frac{%
e^{-m_{n}r}}{m_{n}rw^{\prime }(m_{n})}\int du^{\prime }y_{2}(u^{\prime
},m_{n})du^{\prime } 
\end{equation}
where $m_{n}$ and $y_{2}(u^{\prime },m_{n})$ are the $n$ eigenvalue and
eigenfunction of eq.(\ref{eq:homeq}). Note that the integral in $u^{\prime }$
is well defined for the eigenfunctions. Approximate values of the
eigenvalues and eigenfunctions are obtained in the Appendix.

At long distances the function decays exponentially with a screening length
given by 
\begin{equation}
L_{S}=\frac{R^{2}}{U_{T} m_{gl.}} 
\end{equation}
where $m_{gl}\approx 3.4$ is the lowest eigenvalue. From the boundary point
of view this was expected, since at large temperatures the theory is reduced
to one less dimension. The mass gap in this dimensionally reduced theory is $%
m_{gl.}$ which in the original theory is interpreted as a screening mass.

Hence we find that the screening length is determined by the glueball masses
and is of order $1/T=R^2/U_T$.

\subsection{Confinement and flux tubes in non supersymmetric YM}

The metric (\ref{eq:genbhmetric}) can also be used to describe
nonsupersymmetric QCD$_{p}$, \cite{Witten:1998zw}. To achieve this in the
metric (\ref{eq:genbhmetric}) we must go first to Euclidean time $%
t\rightarrow \tau $ and then go back to Minkowski signature but through $%
x_{p}\rightarrow t$. To avoid a conical singularity at $U=U_{T}$, $\tau $
must be periodic with period $\pi /T=\pi R^{2}/U_{T}$ . As a result the
boundary theory is a SYM theory compactified on a circle and with coupling
constant $g_{YM}^{2}T$, where $g_{YM}^{2}$ is the coupling constant of the
higher dimensional theory.

The potential between two quarks in the boundary is described by the minimal
action of a string going from one quark to another and extending in the
bulk. When the distance between the two charges is very large, the string
hangs down to $U=U_{T}$ where the metric is approximately flat and the
energy is proportional to the separation length giving confinement. This
string is a source for the dilaton and so in the boundary is seen as a non
vanishing $\langle F^{2}\rangle $ which will be concentrated along a flux
tube connecting the two quarks.

The tension of the flux tube was calculated in \cite{Brandhuber:1998er}
using the picture of a hanging string with the result 
\begin{equation}
\sigma _{p}=\frac{1}{2\pi }\left( \frac{U_{T}}{R}\right) ^{\frac{7-p}{2}}
\label{eq:spanning}
\end{equation}
This can be obtained also in a different way. The free energy $A(\beta
,\lambda )$ and the gluon condensate are related by 
\begin{equation}
-\lambda \frac{\partial _{\lambda }A}{\partial \lambda }=\frac{1}{4\lambda }%
\int d^{d}x\langle F^{2}\rangle 
\end{equation}
where we defined $\lambda =g_{YM}^{2}$. In this case the gluon condensate is
known from the boundary value of the dilaton and so the free energy can be
obtained up to a constant as 
\begin{equation}
A=-\int \frac{d\lambda }{\lambda }\int d^{d}x\frac{1}{4\lambda }\langle
F^{2}\rangle . 
\end{equation}
To proceed, we first need the dilaton equation 
\begin{equation}
\frac{1}{2\kappa _{10}^{2}}\partial _{\mu }\sqrt{-g}g^{\mu \nu }e^{-2\bar{\phi}
}\partial _{\nu }\widehat{\phi} =\frac{1}{4\pi \alpha ^{\prime }}\sqrt{-g_{2}}\delta .
\label{eq:dilateq}
\end{equation}
Let us integrate over the $8-p$ dimensions of the internal sphere and the $%
p-2$ transverse directions to the flux tube, i.e. all directions transverse
to the string except the radial $U$ and the angular $\tau $. We put $\chi
=\int d^{6}x\phi $ and obtain 
\begin{equation}
\frac{1}{2\kappa _{10}^{2}}\partial _{U}\sqrt{-g}g^{UU}e^{-2\phi }\partial
_{U}\chi =\frac{1}{4\pi }\left( \frac{U}{R}\right) ^{\frac{7-p}{2}}\delta
^{(2)}\left( U-U_{T}\right) 
\end{equation}
Integrating over the $(U,\tau )$ plane then gives rise to a boundary term at 
$U\rightarrow \infty $. Note that the origin of the radial $U$ coordinate at 
$U=U_{T}$ (the horizon of the Euclidean black hole) is a regular point since
the period $T$ has been chosen in an appropriate way. We find 
\begin{equation}
\frac{\Omega _{8-p}}{2\kappa _{10}^{2}}\int d^{p-2}xd\tau \sqrt{-g}%
g^{UU}e^{-2\bar{\phi}}\partial _{U}\widehat{\phi} =\frac{1}{4\pi }\left( \frac{U_{T}}{R}%
\right) ^{\frac{7-p}{2}}. 
\end{equation}
Note that the leading contribution to $\phi $ is constant over the 
S$^{8-p}$ near the boundary. (In the case p=3 one has that the Kaluza-Klein states on
the sphere correspond to states with higher conformal dimension). Using the
previous results we finally find that the integrated gluon condensate is
given by 
\begin{equation}
\int d^{p-2}xd\tau \frac{1}{4g_{YM}^{2}}F^{2}=\frac{1}{4g_{YM}^{2}T}\int
d^{p-2}xF^{2}=\frac{1}{4\pi }\left( \frac{U_{T}}{R}\right) ^{\frac{7-p}{2}}.
\label{eq:intft}
\end{equation}
We must now relate the gluon condensate to the tension of the flux tube. In
the perturbative case the gluon condensate is proportional to $\lambda $ and
the energy and gluon condensate are equal. The gluon condensate calculated
above is, however, given by $\frac{1}{4\lambda }\langle F^{2}\rangle \sim 
\sqrt{\lambda }$. The reasoning is identically the same as in the case of (%
\ref{eq:kalla}). The result is a free energy that is twice as large as the
gluon condensate\footnote{For a D-string the factor is $-2$.}. This implies 
that (\ref{eq:intft}) correctly
reproduces the tension (\ref{eq:spanning}). More generally we see that, for
massless fields, the total energy in the boundary theory is obtained by
integrating the source over the bulk.

In fact $\langle F^{2}\rangle $ can also be computed as a function of the
position and that gives the profile of the string. We first will look at the
case of a $2+1$ dimensional theory and then at the $3+1$ case.

When $p=3$, the metric is given by 
\begin{eqnarray}
ds^{2} &=&\frac{R^{2}}{U^{2}f(U)}dU^{2}+\frac{U^{2}f(U)}{R^{2}}d\tau ^{2}+%
\frac{U^{2}}{R^{2}}(dx_{1}^{2}+dx_{2}^{2}-dt^{2})+R^{2}d\Omega _{5}^{2} \\
f(U) &=&1-\left( \frac{U_{T}}{U}\right) ^{4}
\end{eqnarray}
where we have let $x_{3}$ become our new time coordinate $t$. A difference
with respect to the non-confining case is that at $U=U_{T}$ there are
geodesics of constant $U$. In fact in that region the metric is
approximately a product metric of an $R^{2+1}$ parameterized by $x_{1,2},t$,
a five sphere and a plane parameterized by $U,\tau $ with $\sqrt{U-U_{T}}$ a
radial coordinate and $\tau $ an angle. From now on we will consider fields
that are independent of the coordinates of the sphere.

A particle sitting at the center $U=U_{T}$ of the (Euclidean) black hole
will remain at rest. Furthermore, a freely falling source will eventually be
trapped around the $R^{2+1}$ at the center and henceforth represent a stable
configuration. This is the reason for the existence of a mass gap and stable
glue balls as well as of confinement. If the particle is a source for the
dilaton, then it will induce in the boundary an expectation value of $%
\langle F^{2}\rangle $ which will look like a stable blob and be interpreted
by the boundary observer as some kind of a glueball state. Consider now the
more physical case of a string sitting at the center of the (euclidean)
black hole and extended for example in the $x_{1}$ direction. In the
boundary it will give rise to a flux tube with a shape that we proceed to
calculate.

First the dilaton field $\phi$ produced by the string should be obtained.
The field is independent of $t$ and $x_1$ since the source is static and
extended in $x_1$ and is also independent of $\tau $ since the string is at
the center of the $\rho =\sqrt{U-U_{T}},\tau $ plane. The equation (\ref
{eq:dilateq}) for $\phi $ is: 
\begin{equation}
\partial _{U}\left( U^{5}f(U)\partial _{U}\phi(U,x_2)\right) +R^{4}U\partial
_{22}\phi(U,x_2)=\frac{2\kappa_{10}^2}{4\pi }\frac{U_T^2}{R^2\Omega_5}
\delta (U-U_{T})\delta(x_2)  \label{eq:propeqebh}
\end{equation}
where $\Omega_5$ is the volume of the unit five-sphere. Rescaling as before $%
U\rightarrow U_{T}u$, $t\rightarrow tR^{2}/U_{T}$ and $x\rightarrow
xR^{2}/U_{T}$, the equation simplifies to 
\begin{equation}
(u^{5}-u)\partial _{uu}\phi+(5u^{4}-1)\partial _{u}\phi
-uk^{2}\phi=\frac{2\kappa_{10}^2}{4\pi }\frac{1}{U_TR^2\Omega_5}%
\delta (u-1)  \label{eq:2p1eq}
\end{equation}
where the Fourier transform 
\begin{equation}
\phi(u,k)=\int dke^{ik}{x_2}\phi(u,x_2) 
\end{equation}
was introduced. The homogeneous equation is the same as (\ref{eq:homeq}). In
terms of its solution the field can be written as 
\begin{equation}
\phi(u,k)=Ay_{1}(u,k) 
\end{equation}
The solution $y_{1}(u,k)$ behaves as $\ln (u-1)$ when $u\rightarrow 1$ as is
needed to match the delta function on the right hand side. In fact we need 
\begin{equation}
\phi(u,k)\approx _{u\rightarrow 1}\frac{1}{4} \frac{2\kappa_{10}^2}{%
4\pi }\frac{1}{U_TR^2\Omega_5}\ln (u-1) 
\end{equation}
On the other hand we have 
\begin{eqnarray}
y_{1}(u,k) &=&\alpha y_{2}(u,k)+\beta \tilde{y}_{2}(u,k) \\
\beta &=&-\frac{y_{1}y_{2}^{\prime }-y_{2}y_{1}^{\prime }}{y_{2}\tilde{y}%
_{2}^{\prime }-y_{2}^{\prime }\tilde{y}_{2}}=-\frac{W(y_{1},y_{2})}{W(y_{2},%
\tilde{y}_{2})}=-\frac{w(k)}{4}
\end{eqnarray}
where $w(k)$ is the function defined in (\ref{eq:defwk}).

The constant $A$ can now be obtained since 
\begin{eqnarray}
\phi(u,k) &=&Ay_{1}(u,k)=\alpha y_{2}(u,k)+\beta \tilde{y}%
_{2}(u,k)\approx _{u\rightarrow 1}A\beta \ln (u-1) \\
&\Rightarrow &A=-\frac{2\kappa _{10}^{2}}{4\pi }\frac{1}{U_{T}R^{2}\Omega
_{5}}\frac{1}{w(k)}
\end{eqnarray}
The field created by the static source is then given by 
\begin{equation}
\phi (u,x_{2})=-\frac{2\kappa _{10}^{2}}{4\pi }\frac{1}{U_{T}R^{2}\Omega _{5}%
}\int \frac{dk}{(2\pi )}e^{-ikx}\frac{1}{w(k)}y_{1}(u,k) 
\end{equation}
Using that near the boundary $y_{1}\approx u^{-4}=U_{T}^{4}/U^{4}$ and using
the relation (\ref{eq:F2dilrel}) between the boundary field and the gluon
condensate, it follows that 
\begin{equation}
\frac{1}{4g_{YM}^{2}}\langle F^{2}\rangle =\frac{4U_{T}^{3}}{4\pi R^{4}}\int 
\frac{dk}{(2\pi )}e^{-ikx_{2}}\frac{1}{w(k)} 
\end{equation}
This will be seen by the boundary observer as an infinite string of finite
width. In the case of two charges separated by a large distance the string
will join both charges. As a check, the integral of this profile over the
transverse direction $x_{2}$ is given by the value of the Fourier transform
at $k=0$. Since $w(0)=4$ (as can be deduced solving the differential
equation exactly for $k=0$) the result (\ref{eq:intft}) is reproduced.

The profile of the QCD string is easily seen to decay exponentially at large
distance as $\exp(-m_{\mbox{gl.}}x_2)$ with $m_{\mbox{gl.}}$ the lowest
glueball mass. An approximation to the profile can be obtained using the
function $w(k)$ computed in the Appendix and is 
\begin{eqnarray}
\frac{1}{4g_{YM}^2}\langle F^2\rangle &\approx& \frac{4U_T^3}{4\pi R^4} \int 
\frac{dk}{(2\pi)} \frac{\pi\left(k^2+\left(\frac{\pi}{2\alpha}%
\right)^2\right)}{16\sqrt{2}} \frac{1}{\cosh(\alpha k)} e^{-ikx_2} \\
&=& \frac{\pi^2}{2^{\frac{15}{2}} } \frac{U_T^3}{R^4\alpha^3} \frac{1}{%
\cosh^3\left(\frac{U_Tx_2}{2\alpha R^2}\right)}
\end{eqnarray}
where in the last line we return to the original variable $x_2\rightarrow
U_Tx_2/R^2$ and $\alpha$ is the constant defined in the Appendix as 
\begin{equation}
\alpha= \frac{1}{4\sqrt{2\pi }}\left( \Gamma (\frac{1}{4})\right) ^{2} 
\end{equation}
This profile is depicted in fig.(\ref{fig:strings}).

For the $p=4$ case the metric and expectation value of the dilaton are given
by 
\begin{eqnarray}
ds^{2} &=&\left( \frac{R}{U}\right) ^{\frac{3}{2}}\frac{dU^{2}}{f(U)}+\left( 
\frac{U}{R}\right) ^{\frac{3}{2}}f(U)d\tau ^{2}+\left( \frac{U}{R}\right) ^{%
\frac{3}{2}}(\sum_{i=1}^{3}dx_{i}^{2}-dt^{2})+R^{\frac{3}{2}}U^{\frac{1}{2}%
}d\Omega _{4}^{2} \nonumber \\
e^{\bar{\phi}} &=&g_{s}\left( \frac{U}{R}\right) ^{\frac{3}{4}} \\
f(U) &=&1-\frac{U_{T}^{3}}{U^{3}} \nonumber
\end{eqnarray}
Considering a string situated at $U=U_{T}$ and extending along $x_{3}$,$t$
as a source, the equation for the massless scalar $\widehat{\phi }$ can then
be written as: 
\begin{equation}
\partial _{U}\left( U^{4}\left( 1-\frac{U_{T}^{3}}{U^{3}}\right) \partial
_{U}\widehat{\phi }\right) +UR^{3}\partial _{ii}\widehat{\phi }=\frac{%
g_{s}^{2}R^{\frac{3}{2}}}{4\pi \Omega _{4}}2\kappa _{10}^{2}\sqrt{%
-g_{tt}g_{33}}\delta ^{(2)}(x_{1,2})\delta (U-U_{T})
\end{equation}
where $\Omega _{4}$ is the volume of the $4$-sphere. Rescaling the variables
as $U=U_{T}u$ and $x_{i}\rightarrow x_{i}R^{\frac{3}{2}}/\sqrt{U_{T}}$ the
equation reduces to 
\begin{equation}
(u^{4}-u)\partial _{uu}\widehat{\phi }+(4u^{3}-1)\partial _{u}\widehat{\phi }%
-uk^{2}\widehat{\phi }=\frac{2\kappa _{10}^{2}}{4\pi \Omega _{4}}\frac{%
g_{s}^{2}}{R^{3}}\frac{1}{\sqrt{U_{T}}}\delta (u-1)
\end{equation}
where the Fourier transform 
\begin{equation}
\widehat{\phi }(u,k)=\int d^{2}ke^{i\vec{k}\vec{x}}\widehat{\phi }%
(x_{1},x_{2},u)
\end{equation}
was introduced. The equation is similar to (\ref{eq:2p1eq}). Two solutions
of the homogeneous eqution can again be defined by 
\begin{equation}
\begin{array}{lcll}
y_{1}(u,k) & = & u^{-3}+\mathcal{O}(u^{-4})\  & u\rightarrow \infty  \\ 
y_{2}(u,k) & = & 1+\mathcal{O}(u-1)\  & u\rightarrow 1
\end{array}
\end{equation}
The Wronskian between these solutions is given by 
\begin{equation}
W(y_{1}(u),y_{2}(u))=y_{1}y_{2}^{\prime }-y_{2}y_{1}^{\prime }=\frac{w(k)}{%
u^{4}-u}
\end{equation}
where $w(k)$ is a function only of $k$ and can be computed as a series
expansion as was done in \cite{Zyskin:1998tg} or in a WKB approximation as
we do in the Appendix. Now the field can be written as 
\begin{equation}
\widehat{\phi }(u,k)=-\frac{2\kappa _{10}^{2}}{4\pi \Omega _{4}}\frac{%
g_{s}^{2}}{R^{3}}\frac{1}{\sqrt{U_{T}}}\frac{1}{w(k)}y_{1}(u,k)
\end{equation}
Expanding this field near the boundary $u\rightarrow \infty $ and using the
relation (\ref{eq:F2dilrel}) with the expectation value of $F^{2}$, the
Fourier transform of the string profile can be obtained as 
\begin{equation}
\frac{1}{4g_{YM}^{2}}\langle F^{2}\rangle (k_{1},k_{2})=\frac{1}{4\pi }%
\left( \frac{U_{T}}{R}\right) ^{\frac{3}{2}}\frac{3}{w(k)}
\end{equation}
Again one can check, using that $w(0)=3$, the integral of this profile
reproduces (\ref{eq:intft}). The string profile is given by 
\begin{eqnarray}
\frac{1}{4g_{YM}^{2}}\langle F^{2}\rangle (k_{1},k_{2}) &=&\frac{1}{4\pi }%
\left( \frac{U_{T}}{R}\right) ^{\frac{3}{2}}\int d^{2}k\frac{3}{w(k)} \\
&=&\frac{1}{4\pi }\left( \frac{U_{T}}{R}\right) ^{\frac{3}{2}}\frac{\pi ^{6}%
}{\alpha ^{5}}\int_{0}^{\infty }\frac{(y^{4}+y^{2})J_{0}\left( \frac{\pi r}{%
\alpha }y\right) }{\sinh \pi y}dy
\end{eqnarray}
where $r^{2}=x_{1}^{2}+x_{2}^{2}$. The last integral cannot be obtained in
closed form, but two expansions valid for large and small $r$ are: 
\begin{eqnarray}
\int_{0}^{\infty }\frac{(y^{4}+y^{2})J_{0}\left( \frac{\pi r}{\alpha }%
y\right) }{\sinh \pi y}dy &=&\frac{2}{\pi }\sum_{n=2}^{\infty
}(-)^{n}n^{2}(n^{2}-1)K_{0}\left( n\frac{\pi r}{\alpha }\right)  \\
&=&\frac{1}{(2\pi )^{5}}\sum_{n=0}^{\infty }c_{n}\frac{r^{2n}}{(4\alpha
)^{2n}}\frac{(-)^{n}}{(n!)^{2}}
\end{eqnarray}
\begin{equation}
c_{n}=2(2^{5+2n}-1)\Gamma (5+2n)\zeta (5+2n)+8\pi ^{2}(2^{3+2n}-1)\Gamma
(3+2n)\zeta (3+2n)
\end{equation}
\mbox{}From the first series it follows that at large distances the profile
decays exponentially as $\exp (-2\pi r/\alpha )$ which is right since $2\pi
/\alpha $ is the mass of the lowest lying glueball in the approximation we
are considering. The profile obtained is depicted in figure \ref{fig:strings}%
.

\FIGURE{\epsfysize=5.0truecm \epsffile{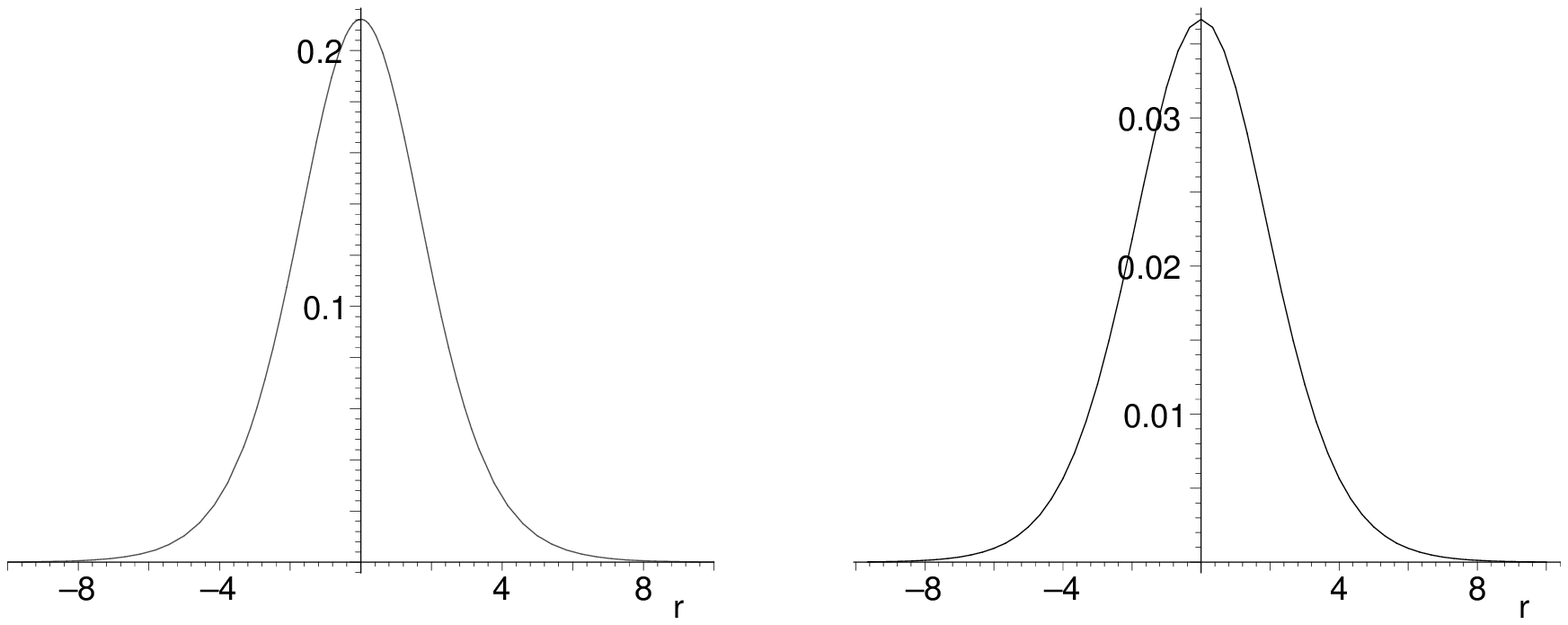}%
\caption{Profile of the $QCD_{2+1}$ (left) and $QCD_{3+1}$ (right) flux
tube. The $r$-axis has units of the corresponding $1/m_{gl}$ and the profile
is normalized so that its integral is $1$.}\label{fig:strings}}

\section{Conclusions}

We have studied various aspects of the bulk/boundary correspondence. First,
we addressed a basic issue about the possible vacuum choices in $AdS$ space.
Often one imposes the system to be in a Poincar\'{e} vacuum state, this
vacuum is however defined with respect to a coordinate system which does not
cover the whole manifold. Hence one might wonder whether it would be more
natural to use the global vacuum state, which is defined with respect to the
global coordinates covering the whole space. However, we showed that for
practical purposes the two vacua can in fact be identified after subtracting
off a constant zero point energy contribution to the vacuum.

We then turned our attention to various probes of the bulk and their
description in the boundary theory. We started by an extension of an example
considered in \cite{Balasub:1998de}, a source in AdS space. Instead of the
quasistatic treatment of \cite{Balasub:1998de}, we took the source to be in
free fall, and evaluated the projection of the source to the boundary by
using retarded propagators instead of static propagators. This results in a
slight modification of the results in \cite{Balasub:1998de}; instead of
looking like a blob with a single scale, its size, the boundary profile
looks like an expanding bubble characterized by two scales which encode the
radial position and the state of motion of the source falling in the bulk.

Static sources are also possible, but in that case they should be kept fixed
by \emph{e.g.} suspending them on a string. The presence of the string is in
turn represented by an additional point charge in the boundary, which
induces a nonzero expectation value for the field strength, $\langle
F^{2}\rangle $. We evaluate this expectation value and find the correct form
for the potential from the point charge.

At non-zero temperatures, the point charge is thermally screened. We
examined this case as well, and evaluated the thermal screening length. It
turned out to be given by glueball masses, of the order of the temperature $%
T $.

We also studied non-supersymmetric, confining $QCD$. In that case, the
center of the (Euclidean) black hole plays a central role. The real world at
the boundary has the character of a shadow of the world at the center of the
Euclidean black hole. A string sitting in the center, appears in the
boundary as a flux tube of finite width with a profile that can be computed.
Pushing this picture a little further, a closed string near the center of
the black hole is mapped into a glueball in the boundary. If the string
state is light, then the center of mass will be spread around $U=U_{T}$
which is the case studied in \cite{Csaki:1998qr}. One can also consider a
qualitative picture of scattering between such 
glueballs\footnote{We thank Bo Sundborg for a discussion on this point.}. 
If the bound states scatter at high energy, two distinct regimes appear. At small
scattering angle the momentum transfer is small and so the particles in
intermediate channels are the lowest lying glueballs. In that domain, when
states far from the $U=U_{T}$ do not contribute, one expects a Regge
behavior from the string scattering amplitudes which corresponds to the
expected Regge behavior of glueball scattering in the boundary. At finite
angles the momentum transfer is large and the region in which the bulk
metric is $AdS_{5}$ is expected to dominate. In the boundary that should
correspond to the parton region in $QCD$, but unfortunately the theory
studied at high energies is not $QCD$ but $N=4$ SYM, which is precisely what
string theory in $AdS_{5}$ describes according to Maldacena's conjecture.

\acknowledgments

We would like to thank Ingemar Bengtsson, Per Kraus, Jorma Louko, and
Bo Sundborg for discussions. The work
of U.D. was supported by Swedish Natural Science Research Council (NFR) and
that of M.K. by The Swedish Foundation for International Cooperation in
Research and Higher Education (STINT).

\section*{Appendix}

In this Appendix we will compute, using a WKB approximation \footnote{%
while this paper was being written ref. \cite{Minahan:1998tm} appear which
has considerable overlap with what follows.} the function $w(k)$ defined in
eq.(\ref{eq:defwk}). To obtain a WKB solution of eq.(\ref{eq:homeq}) it is
better to work with imaginary $k$ and then analytically continue to the real
axis. In that case, the equation (\ref{eq:homeq}) can be recast in the form
of a Schr\"{o}dinger equation for a zero-energy eigenstate \cite
{Csaki:1998qr}: 
\begin{eqnarray}
-\chi ^{\prime \prime }(x)+V(x)\chi &=&0 \\
V(x) &=&-\frac{k^{2}}{4x(x^{2}-1)}+\frac{3x^{4}-6x^{2}-1}{4x^{2}(x^{2}-1)^{2}%
}
\end{eqnarray}
where $x=u^{2}$ and $\chi (x)=\sqrt{x(x^{2}-1)}y(\sqrt{x})$. There are two
regions where the solutions of interest can be expressed in terms of Bessel
functions : 
\begin{equation}
\begin{array}{lcll}
\chi _{1}(x,k) & = & \sqrt{x(x^{2}-1)}y_{1}(\sqrt{x},k)\approx \frac{8\sqrt{x%
}}{k^{2}}J_{2}(\frac{k}{\sqrt{x}}) & \mbox{if }x\gg 1 \\ 
\chi _{2}(x,k) & = & \sqrt{x(x^{2}-1)}y_{2}(\sqrt{x},k)\approx \sqrt{2(x-1)}%
J_{0}(k\sqrt{\frac{x-1}{2}}) & \mbox{if}(x-1)\ll 1
\end{array}
\end{equation}
In the region $(x-1)\gg 1/k^{2}$, $x\ll k^{2}/3$ the potential is
approximately given by 
\begin{equation}
|V(x)|\approx \bar{V}(x)=\frac{k^{2}}{4x(x^{2}-1)}\ . 
\end{equation}
For large $k$ we can use a WKB approximation: 
\begin{equation}
\chi (x)\approx \frac{1}{|\bar{V}|^{\frac{1}{4}}}\left( Ae^{i\int_{x}^{a}%
\sqrt{\bar{V}(x)}}+Be^{-i\int_{x}^{a}\sqrt{\bar{V}(x)}}\right) 
\end{equation}
The strategy now is clear. For large $k$ there is an overlap between the
region $x-1\gg 1/k^{2}$ and $x-1\ll 1$, so we can compute the coefficients $%
A_{2}$ $B_{2}$ corresponding to the function $\chi _{2}$. There is also an
overlap between the regions $x\gg 1$ and $x\ll k^{2}/3$, so we can compute
the coefficients $A_{1}$, $B_{1}$ corresponding to $\chi _{1}$. Finally, the
Wronskian between the two WKB solutions can be computed. The coefficients
are given by 
\begin{equation}
\begin{array}{lclcl}
A_{1} & = & B_{1}^{\ast } & = & \frac{4}{\sqrt{\pi }k^{2}}e^{-\frac{5}{4}%
i\pi +i\frac{k}{\sqrt{a}}}, \\ 
A_{2} & = & B_{2}^{\ast } & = & \frac{1}{\sqrt{2\pi }}e^{\frac{1}{4}i\pi -ik%
\sqrt{\frac{b-1}{2}}+i\int_{b}^{a}\sqrt{\bar{V}(x)}}
\end{array}
\end{equation}
where we assume that the two points $a$, $b$ are such that $1\gg a-1\gg
1/k^{2}$ and $1\ll b\ll k^{2}/3$. The Wronskian can now be obtained as 
\begin{eqnarray}
W(\chi _{1},\chi _{2}) &=&\frac{w(k)}{2}=2i(A_{1}B_{2}-A_{2}B_{1}) \\
&=&-\frac{8\sqrt{2}}{\pi k^{2}}\cos \left( \frac{k}{2}\int_{a}^{b}\frac{dx}{%
\sqrt{x(x^{2}-1)}}+\frac{k}{\sqrt{a}}+k\sqrt{\frac{b-1}{2}}\right) \ .
\end{eqnarray}
The argument of the cosine is approximately independent of $a$ and $b$, so
we can take $b=1$, $a=\infty $. The result is 
\begin{eqnarray}
w(k) &=&-\frac{16\sqrt{2}}{\pi k^{2}}\cos (\alpha k) \\
\alpha &=&\frac{1}{2}\int_{1}^{\infty }\frac{dx}{\sqrt{x(x^{2}-1)}}=\frac{1}{%
4\sqrt{2\pi }}\left( \Gamma (\frac{1}{4})\right) ^{2}
\end{eqnarray}
\mbox{}From here the masses of the $0^{++}$ glueballs follow immediately as 
\begin{equation}
m=\frac{\pi }{\alpha }(n+\frac{1}{2}) 
\end{equation}
The obtained values are in good agreement\footnote{%
As it turns out, the values of \cite{Minahan:1998tm} are even better since
there a next-to-leading order was considered} with the numerical ones except
for the fact that there is an extra eigenvalue at $n=0$. Of course this is
at small $k$, i.e. outside the validity of the approximation. However a
simple modification of the Wronskian cures this problem and gives a good
approximation in the full range $k=0\rightarrow \infty $ for the function $%
w(k)$ obtained numerically. The modification is to write 
\begin{equation}
w(k)=-\frac{16\sqrt{2}}{\pi \left( k^{2}-\left( \frac{\pi }{2\alpha }\right)
^{2}\right) }\cos (\alpha k) 
\end{equation}
In this approximation, $w(0)\approx 5.017$ to be compared with the exact
value $w(0)=4$. Analytically continuing $k\rightarrow ik$ we obtain the
desired function 
\begin{equation}
w(k)=\frac{16\sqrt{2}}{\pi \left( k^{2}+\left( \frac{\pi }{2\alpha }\right)
^{2}\right) }\cosh (\alpha k) 
\end{equation}

The other case studied in the text corresponds to the homogeneous equation 
\begin{equation}
(u^{4}-u)y^{\prime \prime }(u)+(4u^{3}-1)y^{\prime }(u)-uk^{2}y(u)=0 
\end{equation}
The quantity of interest is the Wronskian between the two solutions defined
by 
\begin{equation}
\begin{array}{lcll}
y_{1}(u,k) & = & \frac{1}{u^{3}}+\mathcal{O}(u^{-4})\  & u\rightarrow \infty
\\ 
y_{2}(u,k) & = & 1+\mathcal{O}(u-1)\  & u\rightarrow 1
\end{array}
\end{equation}
To perform a WKB analysis we change variables $x=\sqrt{u}$ and $\chi =\sqrt{%
x^{7}-x}y$. The resulting equation is of the Schr\"{o}dinger type 
\begin{eqnarray}
-\chi ^{\prime \prime }(x)+V(x)\chi &=&0 \\
V(x) &=&\frac{1}{4}\frac{35x^{12}-70x^{6}-1}{x^{2}(x^{6}-1)^{2}}-\frac{%
4x^{3}k^{2}}{x^{7}-x}
\end{eqnarray}
The WKB analysis is completely similar to the previous case. The result is 
\begin{eqnarray}
W(y_{1}(u),y_{2}(u)) &=&\frac{w(k)}{u^{4}-u} \\
w(k) &=&-\frac{6\sqrt{3}}{\pi k^{3}}\sin (\alpha k)
\end{eqnarray}
with 
\begin{equation}
\alpha =\int_{1}^{\infty }\frac{dy}{\sqrt{y^{3}-1}}=\frac{\sqrt{\pi }}{3}%
\frac{\Gamma (\frac{1}{6})}{\Gamma (\frac{2}{3})} 
\end{equation}
There is also a pair of extra poles at $k=\pm \pi /\alpha $. This can be
cured replacing the Wronskian by 
\begin{equation}
w(k)=-\frac{6\sqrt{3}}{\pi k}\frac{1}{k^{2}-\frac{\pi ^{2}}{\alpha ^{2}}}%
\sin (\alpha k) 
\end{equation}
\mbox{}From where we read the glueball masses as 
\begin{equation}
m_{n}=\frac{\pi }{\alpha }n,\ \ n=2,3,\ldots 
\end{equation}
in good agreement with the results of \cite{Csaki:1998qr}. Analytically
continuing to $k\rightarrow ik$ the desired Wronskian is 
\begin{equation}
w(k)=\frac{6\sqrt{3}}{\pi k}\frac{1}{k^{2}+\frac{\pi ^{2}}{\alpha ^{2}}}%
\sinh (\alpha k) 
\end{equation}

\providecommand{\href}[2]{#2}\begingroup\raggedright\endgroup


\begin{thebibliography}{10}

\bibitem{Maldacena:1997re}
J.~Maldacena, {\it The large N limit of superconformal field theories and
  supergravity},  {\em Adv. Theor. Math. Phys.} {\bf 2} (1998) 231,
  [\href{http://xxx.lanl.gov/abs/hep-th/9711200}{{\tt hep-th/9711200}}].

\bibitem{Polyakov:1998ju}
A.~M. Polyakov, {\it The wall of the cave},
  \href{http://xxx.lanl.gov/abs/hep-th/9809057}{{\tt hep-th/9809057}}.

\bibitem{Polyakov:1997tj}
A.~M. Polyakov, {\it String theory and quark confinement},  {\em Nucl. Phys.
  Proc. Suppl.} {\bf 68} (1998) 1,
  [\href{http://xxx.lanl.gov/abs/hep-th/9711002}{{\tt hep-th/9711002}}].

\bibitem{Gubser:1998bc}
S.~S. Gubser, I.~R. Klebanov, and A.~M. Polyakov, {\it Gauge theory correlators
  from noncritical string theory},  {\em Phys. Lett.} {\bf B428} (1998) 105,
  [\href{http://xxx.lanl.gov/abs/hep-th/9802109}{{\tt hep-th/9802109}}].

\bibitem{Witten:1998qj}
E.~Witten, {\it Anti-de Sitter space and holography},  {\em Adv. Theor. Math.
  Phys.} {\bf 2} (1998) 253,
  [\href{http://xxx.lanl.gov/abs/hep-th/9802150}{{\tt hep-th/9802150}}].

\bibitem{Rey:1998ik}
S.-J. Rey and J.~Yee, {\it Macroscopic strings as heavy quarks in large N gauge
  theory and anti-de Sitter supergravity},
  \href{http://xxx.lanl.gov/abs/hep-th/9803001}{{\tt hep-th/9803001}}.

\bibitem{Maldacena:1998im}
J.~Maldacena, {\it Wilson loops in large N field theories},  {\em Phys. Rev.
  Lett.} {\bf 80} (1998) 4859,
  [\href{http://xxx.lanl.gov/abs/hep-th/9803002}{{\tt hep-th/9803002}}].

\bibitem{Brandhuber:1998er}
A.~Brandhuber, N.~Itzhaki, J.~Sonnenschein, and S.~Yankielowicz, {\it Wilson
  loops, confinement, and phase transitions in large N gauge theories from
  supergravity},  {\em J. High Energy Phys.} {\bf 06} (1998) 001,
  [\href{http://xxx.lanl.gov/abs/hep-th/9803263}{{\tt hep-th/9803263}}].

\bibitem{Rey:1998bq}
S.-J. Rey, S.~Theisen, and J.-T. Yee, {\it Wilson-polyakov loop at finite
  temperature in large N gauge theory and anti-de Sitter supergravity},  {\em
  Nucl. Phys.} {\bf B527} (1998) 171,
  [\href{http://xxx.lanl.gov/abs/hep-th/9803135}{{\tt hep-th/9803135}}].

\bibitem{Minahan:1998xb}
J.~A. Minahan, {\it Quark - monopole potentials in large N SuperYang-Mills},
  {\em Adv. Theor. Math. Phys.} {\bf 2} (1998) 559,
  [\href{http://xxx.lanl.gov/abs/hep-th/9803111}{{\tt hep-th/9803111}}].

\bibitem{Brandhuber:1998bs}
A.~Brandhuber, N.~Itzhaki, J.~Sonnenschein, and S.~Yankielowicz, {\it Wilson
  loops in the large N limit at finite temperature},  {\em Phys. Lett.} {\bf
  B434} (1998) 36--40, [\href{http://xxx.lanl.gov/abs/hep-th/9803137}{{\tt
  hep-th/9803137}}].

\bibitem{Danielsson:1998br}
U.~H. Danielsson and A.~P. Polychronakos, {\it Quarks, monopoles and dyons at
  large N},  {\em Phys. Lett.} {\bf B434} (1998) 294--302,
  [\href{http://xxx.lanl.gov/abs/hep-th/9804141}{{\tt hep-th/9804141}}].

\bibitem{Witten:1998zw}
E.~Witten, {\it Anti-de Sitter space, thermal phase transition, and confinement
  in gauge theories},  {\em Adv. Theor. Math. Phys.} {\bf 2} (1998) 505,
  [\href{http://xxx.lanl.gov/abs/hep-th/9803131}{{\tt hep-th/9803131}}].

\bibitem{Gross:1998gk}
D.~J. Gross and H.~Ooguri, {\it Aspects of large N gauge theory dynamics as
  seen by string theory},  {\em Phys. Rev.} {\bf D58} (1998) 106002,
  [\href{http://xxx.lanl.gov/abs/hep-th/9805129}{{\tt hep-th/9805129}}].

\bibitem{Ooguri:1998hq}
H.~Ooguri, H.~Robins, and J.~Tannenhauser, {\it Glueballs and their
  Kaluza-Klein cousins},  {\em Phys. Lett.} {\bf B437} (1998) 77,
  [\href{http://xxx.lanl.gov/abs/hep-th/9806171}{{\tt hep-th/9806171}}].

\bibitem{deMelloKoch:1998qs}
R.~de~Mello~Koch, A.~Jevicki, M.~Mihailescu, and J.~P. Nunes, {\it Evaluation
  of glueball masses from supergravity},  {\em Phys. Rev.} {\bf D58} (1998)
  105009, [\href{http://xxx.lanl.gov/abs/hep-th/9806125}{{\tt
  hep-th/9806125}}].

\bibitem{Csaki:1998qr}
C.~Csaki, H.~Ooguri, Y.~Oz, and J.~Terning, {\it Glueball mass spectrum from
  supergravity},  \href{http://xxx.lanl.gov/abs/hep-th/9806021}{{\tt
  hep-th/9806021}}.

\bibitem{Hashimoto:1998if}
A.~Hashimoto and Y.~Oz, {\it Aspects of QCD dynamics from string theory},
  \href{http://xxx.lanl.gov/abs/hep-th/9809106}{{\tt hep-th/9809106}}.

\bibitem{Russo:1998mm}
J.~G. Russo, {\it New compactifications of supergravities and large N QCD},
  \href{http://xxx.lanl.gov/abs/hep-th/9808117}{{\tt hep-th/9808117}}.

\bibitem{Csaki:1998cb}
C.~Csaki, Y.~Oz, J.~Russo, and J.~Terning, {\it Large N QCD from rotating
  branes},  \href{http://xxx.lanl.gov/abs/hep-th/9810186}{{\tt
  hep-th/9810186}}.

\bibitem{Balasub:1998sn}
V.~Balasubramanian, P.~Kraus, and A.~Lawrence, {\it Bulk vs. boundary dynamics
  in anti-de Sitter space-time},
  \href{http://xxx.lanl.gov/abs/hep-th/9805171}{{\tt hep-th/9805171}}.

\bibitem{Balasub:1998de}
V.~Balasubramanian, P.~Kraus, A.~Lawrence, and S.~P. Trivedi, {\it Holographic
  probes of anti-de Sitter space-times},
  \href{http://xxx.lanl.gov/abs/hep-th/9808017}{{\tt hep-th/9808017}}.

\bibitem{Maldacena:1998bw}
J.~Maldacena and A.~Strominger, {\it AdS(3) black holes and a stringy exclusion
  principle},  \href{http://xxx.lanl.gov/abs/hep-th/9804085}{{\tt
  hep-th/9804085}}.

\bibitem{Keski-Vakkuri:1998nw}
E.~Keski-Vakkuri, {\it Bulk and boundary dynamics in BTZ black holes},
  \href{http://xxx.lanl.gov/abs/hep-th/9808037}{{\tt hep-th/9808037}}.

\bibitem{Muller-Kirsten:1998mt}
H.~J.~W. Muller-Kirsten, N.~Ohta, and J.-G. Zhou, {\it AdS(3) / CFT
  correspondence, poincare vacuum state and grey body factors in BTZ black
  holes},  \href{http://xxx.lanl.gov/abs/hep-th/9809193}{{\tt hep-th/9809193}}.

\bibitem{Ohta:1998xh}
N.~Ohta and J.-G. Zhou, {\it Thermalization of Poincare vacuum state and
  fermion emission from AdS(3) black holes in bulk boundary correspondence},
  \href{http://xxx.lanl.gov/abs/hep-th/9811057}{{\tt hep-th/9811057}}.

\bibitem{Banados:1992wn}
M.~Banados, C.~Teitelboim, and J.~Zanelli, {\it The black hole in
  three-dimensional space-time},  {\em Phys. Rev. Lett.} {\bf 69} (1992)
  1849--1851, [\href{http://xxx.lanl.gov/abs/hep-th/9204099}{{\tt
  hep-th/9204099}}].

\bibitem{Banados:1993gq}
M.~Banados, M.~Henneaux, C.~Teitelboim, and J.~Zanelli, {\it Geometry of the
  (2+1) black hole},  {\em Phys. Rev.} {\bf D48} (1993) 1506--1525,
  [\href{http://xxx.lanl.gov/abs/gr-qc/9302012}{{\tt gr-qc/9302012}}].

\bibitem{Natsuume:1996ij}
M.~Natsuume and Y.~Satoh, {\it String theory on three-dimensional black holes},
   {\em Int. J. Mod. Phys.} {\bf A13} (1998) 1229,
  [\href{http://xxx.lanl.gov/abs/hep-th/9611041}{{\tt hep-th/9611041}}].

\bibitem{Satoh:1997xf}
Y.~Satoh, {\it Study of three-dimensional quantum black holes},
  \href{http://xxx.lanl.gov/abs/hep-th/9705209}{{\tt hep-th/9705209}}.

\bibitem{Vil}
N.Ja.Vilenkin and A.V.Klimyk, {\em Representations of Lie groups and special
  functions}.
\newblock Kluwer, 1991.

\bibitem{Bombelli:1986rw}
L.~Bombelli, R.~K. Koul, J.~Lee, and R.~D. Sorkin, {\it A quantum source of
  entropy for black holes},  {\em Phys. Rev.} {\bf D34} (1986) 373.

\bibitem{Srednicki:1993im}
M.~Srednicki, {\it Entropy and area},  {\em Phys. Rev. Lett.} {\bf 71} (1993)
  666--669, [\href{http://xxx.lanl.gov/abs/hep-th/9303048}{{\tt
  hep-th/9303048}}].

\bibitem{Hol}
C.Holzhey.
\newblock PhD thesis, Princeton University, 1992.
\newblock (unpublished).

\bibitem{Birrell:1982ix}
N.~D. Birrell and P.~C.~W. Davies, {\em Quantum Fields in Curved Space}.
\newblock Cambridge, Uk: Univ. Pr. (1982).

\bibitem{Brown:1986nw}
J.~D. Brown and M.~Henneaux, {\it Central charges in the canonical realization
  of asymptotic symmetries: An example from three-dimensional gravity},  {\em
  Commun. Math. Phys.} {\bf 104} (1986) 207.

\bibitem{Jackiw:1995be}
R.~Jackiw, {\em Diverse topics in theoretical and mathematical physics}.
\newblock Singapore, Singapore: World Scientific (1995).

\bibitem{Deser:1998xb}
S.~Deser and O.~Levin, {\it Mapping Hawking into Unruh thermal properties},
  \href{http://xxx.lanl.gov/abs/hep-th/9809159}{{\tt hep-th/9809159}}.

\bibitem{Breitenlohner:1982jf}
P.~Breitenlohner and D.~Z. Freedman, {\it Stability in gauged extended
  supergravity},  {\em Ann. Phys.} {\bf 144} (1982) 249.

\bibitem{Breitenlohner:1982bm}
P.~Breitenlohner and D.~Z. Freedman, {\it Positive energy in anti-de Sitter
  backgrounds and gauged extended supergravity},  {\em Phys. Lett.} {\bf 115B}
  (1982) 197.

\bibitem{Mezincescu:1985ev}
L.~Mezincescu and P.~K. Townsend, {\it Stability at a local maximum in higher
  dimensional anti-de Sitter space and applications to supergravity},  {\em
  Ann. Phys.} {\bf 160} (1985) 406.

\bibitem{Berenstein:1998ij}
D.~Berenstein, R.~Corrado, W.~Fischler, and J.~Maldacena, {\it The operator
  product expansion for Wilson loops and surfaces in the large N limit},
  \href{http://xxx.lanl.gov/abs/hep-th/9809188}{{\tt hep-th/9809188}}.

\bibitem{Grad}
I.~R. I.S.~Gradshteyn, {\em Table of integrals, series and products}.
\newblock Academic Press: San Diego, fourth~ed., 1992.

\bibitem{Avis:1978yn}
S.~J. Avis, C.~J. Isham, and D.~Storey, {\it Quantum field theory in anti-de
  Sitter space-time},  {\em Phys. Rev.} {\bf D18} (1978) 3565.

\bibitem{Chepelev:1997vx}
I.~Chepelev and A.~A. Tseytlin, {\it Long distance interactions of D-brane
  bound states and longitudinal five-brane in M(atrix) theory},  {\em Phys.
  Rev.} {\bf D56} (1997) 3672--3685,
  [\href{http://xxx.lanl.gov/abs/hep-th/9704127}{{\tt hep-th/9704127}}].

\bibitem{Zyskin:1998tg}
M.~Zyskin, {\it A note on the glueball mass spectrum},
  \href{http://xxx.lanl.gov/abs/hep-th/9806128}{{\tt hep-th/9806128}}.

\bibitem{Minahan:1998tm}
J.~A. Minahan, {\it Glueball mass spectra and other issues for supergravity
  duals of QCD models},  \href{http://xxx.lanl.gov/abs/hep-th/9811156}{{\tt
  hep-th/9811156}}.

\end{thebibliography}
\end{document}